\documentclass[seceq]{ptptex}
\usepackage{graphicx, wrapft, braket}
\input{colordvi.tex}

\title{Parity Violation of Gravitons in the CMB Bispectrum}

\author{Maresuke \textsc{Shiraishi},\footnote{Email: mare@a.phys.nagoya-u.ac.jp} 
Daisuke \textsc{Nitta}\footnote{Email: nitta@a.phys.nagoya-u.ac.jp} 
and Shuichiro \textsc{Yokoyama}\footnote{Email: shu@a.phys.nagoya-u.ac.jp} 
}
\inst{Department of Physics and Astrophysics, Nagoya University, Nagoya 464-8602, Japan}

\abst{
We investigate the cosmic microwave background (CMB)
 bispectra of the intensity (temperature) and polarization modes
induced by the graviton non-Gaussianities,
which arise from the
parity-conserving and parity-violating Weyl cubic terms with
time-dependent coupling.
By considering the time-dependent coupling,
we find that 
even in the exact de Sitter space time, the parity violation still appears in the three-point function 
of the primordial gravitational waves
and could become large.
Through the estimation of the CMB bispectra, we  
demonstrate that the signals generated from the parity-conserving and parity-violating terms appear in completely different
configurations of multipoles. 
For example,
the parity-conserving non-Gaussianity
induces the nonzero CMB temperature bispectrum
in the configuration with 
$\sum_{n=1}^3 \ell_n = {\rm even}$ 
and, while due to the parity-violating non-Gaussianity, the CMB temperature bispectrum also appears for $\sum_{n=1}^3 \ell_n =
{\rm odd}$. This signal is just good evidence
of the parity violation in the non-Gaussianity of primordial
gravitational waves. We find that 
the shape of this non-Gaussianity is similar to the so-called equilateral one
and the amplitudes of these spectra at large scale are roughly estimated as
$|b_{\ell \ell \ell}| \sim \ell^{-4} \times 3.2 \times 10^{-2} \left( {\rm GeV} / \Lambda \right)^2 \left(r / 0.1\right)^4$,
where $\Lambda$ is an energy scale that sets the magnitude of the Weyl
cubic terms (higher derivative corrections) and $r$ is a tensor-to-scalar ratio.
Taking the limit for the nonlinearity parameter of the equilateral type as
$f_{\rm NL}^{\rm eq} < 300$, we can obtain a bound as $\Lambda \gtrsim
3 \times 10^6 {\rm GeV}$, assuming $r=0.1$. 
}

\PTPindex{400, 434, 435, 442, 452}  

\begin{document}
\maketitle

\section{Introduction}


Non-Gaussian features in the cosmological perturbations include detailed 
information on the nature of the early Universe, and 
there have
been many works that attempt to extract them from the bispectrum (three-point function)
of the cosmic microwave background (CMB) anisotropies
(e.g., Refs.~\citen{Komatsu:2001rj, Bartolo:2004if, Babich:2004gb, Komatsu:2010fb}). 
However, most of these discussions are limited in the cases that the
scalar-mode contribution dominates in the non-Gaussianity and also
are based on the assumption of rotational invariance and parity conservation. 

In contrast, there are several studies on the non-Gaussianities of not
only the scalar-mode perturbations but also the vector- and
tensor-mode perturbations \cite{Maldacena:2002vr, Brown:2005kr,
Adshead:2009bz}. 
These sources produce the additional signals on the CMB bispectrum
\cite{Shiraishi:2010kd} and can give a dominant contribution by 
considering such highly non-Gaussian sources as the stochastic magnetic fields \cite{Shiraishi:2011dh}.
Furthermore, even in the CMB bispectrum induced from the scalar-mode non-Gaussianity, 
if the rotational invariance is violated in the non-Gaussianity, 
the characteristic signals appear \cite{Shiraishi:2011ph}. 
Thus, it is very important to clarify these less-noted signals to understand the precise picture of the early Universe. 

Recently, the parity violation in the graviton non-Gaussianities has 
been discussed in Refs.~\citen{Maldacena:2011nz} and \citen{Soda:2011am}. 
Maldacena and
Pimentel first calculated the primordial bispectrum of the gravitons
sourced from parity-even (parity-conserving) and parity-odd (parity-violating) Weyl
cubic terms, namely, $W^3$ and $\widetilde{W}W^2$, respectively, by making use of the spinor
helicity formalism.\cite{Maldacena:2011nz} Soda $\it et \ al.$ proved that the parity-violating
non-Gaussianity of the primordial gravitational waves induced from $\widetilde{W}W^2$ emerges not in the
exact de Sitter space-time but in the quasi de Sitter space-time, and hence, its
amplitude is proportional to a slow-roll parameter.\cite{Soda:2011am} In these studies,
the authors assume that the coupling constant of the Weyl cubic terms is independent of time.

In this paper, we estimate the primordial
non-Gaussianities of gravitons generated from $W^3$ and
$\widetilde{W}W^2$ with the time-dependent coupling parameter~\cite{Weinberg:2008hq}.
We consider the case where the coupling is given by a power of the
conformal time.
We show that in such a model, the parity violation in the non-Gaussianity of the
primordial gravitational waves would not vanish even in the exact de Sitter space-time.
The effects of the parity violation on the CMB power spectrum have been well-studied,
where an attractive result is that the cross-correlation between 
the intensity and $B$-mode polarization is generated \cite{Alexander:2004wk, Saito:2007kt, Gluscevic:2010vv, Sorbo:2011rz}. 
On the other hand, in the CMB bispectrum, owing to the mathematical
property of the spherical harmonic function, 
the parity-even and parity-odd signals should arise from just the
opposite configurations of multipoles \cite{Okamoto:2002ik, Kamionkowski:2010rb}.
Then, we 
formulate and numerically calculate the CMB bispectra induced by these
non-Gaussianities that contain all the correlations between the intensity ($I$)
and polarizations ($E,B$) and
show that 
the signals from $W^3$ (parity-conserving)
appear in the configuration of the multipoles
where those from $\widetilde{W}W^2$ (parity-violating) vanish and vice versa. 

This paper is organized as follows. In the next section, we derive the
primordial bispectrum of gravitons induced by $W^3$ 
and $\widetilde{W}W^2$ with the coupling constant proportional to
the power of the conformal time. In \S\ref{sec:CMB_bis}, we calculate
the CMB bispectra sourced from these non-Gaussianities, analyze their
behavior and find some peculiar signatures of the parity violation. 
The final section is devoted to summary and discussion. In Appendix
\ref{appen:pol_tens}, we explain the detailed calculation of the product
between the polarization tensors and unit vectors.

Throughout this paper, we use $M_{\rm pl} \equiv 1/\sqrt{8 \pi G}$,
where $G$ is the Newton constant and the rule that all the Greek characters and alphabets run from $0$ to $3$ and from $1$ to $3$, respectively. 

\section{Parity-even and parity-odd non-Gaussianity of gravitons}

In this section, we formulate the primordial non-Gaussianity of gravitons
generated from the Weyl cubic terms with the running coupling constant as a
function of a conformal time, $f(\tau)$, whose action is given by
\begin{eqnarray}
S = \int d\tau d^3x \frac{f(\tau)}{\Lambda^2}
\left( \sqrt{-g} W^3  +  \widetilde{W}W^2 \right)~, \label{eq:action}
\end{eqnarray}
with
\begin{eqnarray}
W^3 &\equiv& W^{\alpha \beta}{}_{\gamma \delta} W^{\gamma
 \delta}{}_{\sigma \rho} W^{\sigma \rho}{}_{\alpha \beta}~, \\
\widetilde{W}W^2 &\equiv& \epsilon^{\alpha \beta \mu \nu} 
W_{\mu \nu \gamma \delta} W^{\gamma \delta}{}_{\sigma \rho} 
W^{\sigma \rho}{}_{\alpha \beta}~,
\end{eqnarray}
where $W^{\alpha \beta}{}_{\gamma \delta}$ denotes the Weyl tensor,
$\epsilon^{\alpha \beta \mu \nu}$ is a 4D Levi-Civita tensor
normalized as $\epsilon^{0123} = 1$, and $\Lambda$ is a scale that sets the
value of the higher derivative corrections \cite{Maldacena:2011nz}. 
Note that $W^3$ and $\widetilde{W}W^2$ have the even and odd parities,
respectively. 
In the following discussion, we assume that the coupling constant is given by
\begin{eqnarray}
f(\tau) = \left( \frac{\tau}{\tau_*} \right)^A ~,
\end{eqnarray} 
where $\tau$ is a conformal time.
Here, we have set $f(\tau_*) = 1$.
Such a coupling can be readily realized by considering a dilaton-like coupling in the slow-roll inflation
as discussed in \S\ref{subsec:coupling}.

\subsection{Calculation of the primordial bispectrum}

Here, let us focus on the calculation of the primordial
bispectrum induced by $W^3$ and $\widetilde{W}W^2$ of
Eq.~(\ref{eq:action}) on the exact de Sitter space-time in a more straightforward manner than those of
Refs.~\citen{Maldacena:2011nz} and \citen{Soda:2011am}. 

At first, we consider the tensor perturbations on the Friedmann-Lemaitre-Robertson-Walker metric as
\begin{eqnarray}
ds^2 = a^2 ( -d\tau^2 + e^{\gamma_{ij}} dx^i dx^j)~,
\end{eqnarray}
where $a$ denotes the scale factor and $\gamma_{ij}$ obeys the transverse traceless conditions;
$\gamma_{ii} = \partial \gamma_{ij}/ \partial x^j =0$.  
Up to the second order, even if the action includes the Weyl cubic terms
given by Eq.~(\ref{eq:action}),
the gravitational wave obeys the action as
\cite{Maldacena:2011nz, Soda:2011am}
\begin{eqnarray}
S = \frac{M_{\rm pl}^2}{8} \int d\tau dx^3 a^2(\dot{\gamma}_{ij}
 \dot{\gamma}_{ij} - \gamma_{ij,k} \gamma_{ij,k})~,
\end{eqnarray}
where $~\dot~ \equiv \partial/\partial \tau$ and $_{,i} \equiv \partial/\partial x^i$. 
We expand the gravitational wave with a transverse and traceless
polarization tensor $e^{(\lambda)}_{ij}$ and
the creation and annihilation operators $a^{(\lambda) \dagger},
a^{(\lambda)}$ as 
\begin{eqnarray}
\gamma_{ij}(\mib{x}, \tau) &=& \int \frac{d^3 \mib{k}}{(2\pi)^3}
 \sum_{\lambda = \pm 2} \gamma_{dS}(k,\tau) a^{(\lambda)}_{\mib{k}} e^{(\lambda)}_{ij}(\hat{\mib{k}}) e^{i \mib{k} \cdot \mib{x}} + h.c. \nonumber \\
&=& \int \frac{d^3 \mib{k}}{(2\pi)^3} \sum_{\lambda = \pm 2} 
\gamma^{(\lambda)}(\mib{k}, \tau)
e^{(\lambda)}_{ij}(\hat{\mib{k}}) e^{i \mib{k} \cdot \mib{x}}~, 
\end{eqnarray}
with 
\begin{eqnarray}
\gamma^{(\lambda)}(\mib{k}, \tau) 
&\equiv& \gamma_{dS}(k,\tau) a^{(\lambda)}_{\mib{k}} 
+  \gamma^*_{dS}(k,\tau) a^{(\lambda) \dagger}_{-\mib{k}}~.
\end{eqnarray}
Here, $\lambda \equiv \pm 2$ denotes the helicity of the gravitational wave and we use the polarization tensor satisfying the relations as  
\begin{eqnarray}
e_{ii}^{(\lambda)}(\hat{\mib{k}}) &=& \hat{k}_i e_{ij}^{(\lambda)}(\hat{\mib{k}}) = 0~, \nonumber \\
e_{ij}^{(\lambda) *}(\hat{\mib{k}}) &=& e_{ij}^{(-\lambda)}(\hat{\mib{k}}) = e_{ij}^{(\lambda)}(- \hat{\mib{k}})~, \nonumber \\
e_{ij}^{(\lambda)}(\hat{\mib{k}}) e_{ij}^{(\lambda')}(\hat{\mib{k}}) &=& 2
\delta_{\lambda, -\lambda'}~. \label{eq:pol_tens_rel}
\end{eqnarray} 
The creation and annihilation operators
$a^{(\lambda) \dagger}, a^{(\lambda)}$ obey the relations as
\begin{eqnarray}
a^{(\lambda)}_{\mib{k}} \Ket{0} &=& 0~, \nonumber \\
\left[ a^{(\lambda)}_{\mib{k}}, a^{(\lambda') \dagger}_{\mib{k'}}
\right] &=& (2\pi)^3 \delta(\mib{k} - \mib{k'}) \delta_{\lambda,
\lambda'}~, \label{eq:a}
\end{eqnarray}
where $\ket{0}$ denotes a vacuum eigenstate. 
Then, the mode function of gravitons on the de Sitter space-time
$\gamma_{dS}$ satisfies the field equation as 
\begin{eqnarray}
  \ddot{\gamma}_{dS} - \frac{2}{\tau} \dot{\gamma}_{dS} + k^2 \gamma_{dS}  = 0~,
\end{eqnarray}
and a solution is given by
\begin{eqnarray}
\gamma_{dS} = i \frac{H}{M_{\rm pl} } 
\frac{e^{- i k \tau}}{k^{3/2}} (1 + ik\tau)~,
\label{eq:sol}
\end{eqnarray}
where $H = - (a\tau)^{-1}$ is the Hubble parameter and has a constant value
 in the exact de Sitter space-time.

On the basis of the in-in formalism (see, e.g.,
Refs.~\citen{Maldacena:2002vr} and \citen{Weinberg:2005vy}) and the above results, we calculate the tree-level
bispectrum of gravitons on the late-time limit. According to this
formalism, the expectation value of an operator depending on time in the interaction picture,
$O(t)$, is written as
\begin{eqnarray}
\Braket{O(t)} = \Braket{ 0 |\bar{T} e^{i \int H_{int}(t') dt'} O(t) 
T e^{-i \int H_{int}(t') dt'} | 0 } ~,
\end{eqnarray} 
where $T$ and $\bar{T}$ are respectively time-ordering and anti-time-ordering operators
and  $H_{int}(t)$ is the interaction Hamiltonian.
Applying this
equation, the primordial bispectrum of gravitons at the tree level can be
expressed as 
\begin{eqnarray}
\Braket{\prod_{n=1}^3 \gamma^{(\lambda_n)}(\mib{k_n},\tau)} 
= i \int_{- \infty}^\tau d\tau' \Braket{ 0 | 
\left[: H_{int}(\tau'):, \prod_{n=1}^3
 \gamma^{(\lambda_n)}(\mib{k_n},\tau) \right] | 0 }, \label{eq:in-in_formalism} 
\end{eqnarray} 
where $:~:$ denotes normal product. 

Up to the first order with respect to $\gamma_{ij}$, the nonzero components of the Weyl tensor are written as
\begin{eqnarray}
W^{0i}{}_{0j} &=& \frac{1}{4}(H\tau)^2 \gamma_{ij, \alpha \alpha}~,
 \nonumber \\
W^{ij}{}_{0k} &=& \frac{1}{2}(H\tau)^2(\dot{\gamma}_{ki,j} -
 \dot{\gamma}_{kj,i})~, \nonumber \\
W^{0i}{}_{jk} &=& \frac{1}{2}(H\tau)^2(\dot{\gamma}_{ik,j} -
 \dot{\gamma}_{ij,k})~, \nonumber \\
W^{ij}{}_{kl} &=& \frac{1}{4}(H\tau)^2
(-\delta_{ik} \gamma_{jl,\alpha \alpha} + \delta_{il}
\gamma_{jk,\alpha \alpha} + \delta_{jk} \gamma_{il,\alpha \alpha} -
\delta_{jl} \gamma_{ik,\alpha \alpha} )~, \label{eq:W}
\end{eqnarray}
where $\gamma_{ij, \alpha \alpha} \equiv \ddot{\gamma}_{ij} + \nabla^2
\gamma_{ij}$. Then, $W^3$ and $\widetilde{W}W^2$ respectively reduce to 
\begin{eqnarray}
W^3 &=& W^{ij}{}_{kl} W^{kl}{}_{mn} W^{mn}{}_{ij} + 6 W^{0i}{}_{jk} W^{jk}{}_{lm} W^{lm}{}_{0i} \nonumber \\
&&+ 12 W^{0i}{}_{0j} W^{0j}{}_{kl}  W^{kl}{}_{0i} + 8 W^{0i}{}_{0j}
 W^{0j}{}_{0k}  W^{0k}{}_{0i} ~, \\
\widetilde{W}W^2 
&=& 4 \eta^{ijk} 
\left[ W_{jkpq} 
\left( W^{pq}{}_{lm} W^{lm}{}_{0i} + 2 W^{pq}{}_{0 m}
 W^{0m}{}_{0i} \right) \right. \nonumber \\
&&\qquad \left. + 2 W_{jk0p} \left( W^{0p}{}_{lm} W^{lm}{}_{0i} + 2 W^{0p}{}_{0 m} W^{0m}{}_{0i} \right) \right]~,
\end{eqnarray}
where $\eta^{ijk} \equiv \epsilon^{0ijk}$. 
Using the above expressions and $\int d\tau H_{int} = -S_{int}$,
up to the third order,
the interaction Hamiltonians of $W^3$ and $\widetilde{W}W^2$ are respectively given by
\begin{eqnarray}
H_{W^3} &=& - \int d^3x \Lambda^{-2} 
 (H \tau)^2 \left(\frac{\tau}{\tau_*}\right)^A \nonumber \\
&&\times \frac{1}{4} \gamma_{ij,\alpha \alpha} 
\left[ 
 \gamma_{jk,\beta \beta} \gamma_{ki,\sigma \sigma}
 + 6 \dot{\gamma}_{kl,i}\dot{\gamma}_{kl,j} 
 + 6 \dot{\gamma}_{ik,l}\dot{\gamma}_{jl,k}
 - 12 \dot{\gamma}_{ik,l}\dot{\gamma}_{kl,j}
\right]~, 
\label{eq:Lag_aW3} \\
H_{\widetilde{W}W^2} 
&=& - \int d^3 x 
 \Lambda^{-2} (H \tau)^2 \left(\frac{\tau}{\tau_*}\right)^A \nonumber \\
&&\times
\eta^{ijk}
\left[ \gamma_{kq,\alpha \alpha} 
( - 3 
{\gamma}_{jm,\beta \beta}  \dot{\gamma}_{iq, m}
+ {\gamma}_{mi,\beta \beta } \dot{\gamma}_{mq,j} ) + 4 \dot{\gamma}_{pj,k}
\dot{\gamma}_{pm,l} ( \dot{\gamma}_{il,m} - \dot{\gamma}_{im,l} )
\right]~.
\nonumber \\
\end{eqnarray}
Substituting the above expressions into Eq.~(\ref{eq:in-in_formalism}), using the solution given by Eq.~(\ref{eq:sol}), and considering the
late-time limit as $\tau \rightarrow 0$, 
we can obtain an explicit form of the primordial bispectra: 
\begin{eqnarray}
\Braket{\prod_{n=1}^3 \gamma^{(\lambda_n)}(\mib{k_n})}_{int}
&=&  (2\pi)^3 \delta \left(\sum_{n=1}^3 \mib{k_n}\right)
f^{(r)}_{int}(k_1, k_2, k_3) 
f^{(a)}_{int}(\hat{\mib{k_1}}, \hat{\mib{k_2}}, \hat{\mib{k_3}})~,
\end{eqnarray}
with\footnote{Here, we set $\tau_* < 0$.}
\begin{eqnarray}
f^{(r)}_{W^3} 
&=& 8 \left(\frac{H}{M_{\rm pl}}\right)^6 \left( \frac{H}{\Lambda} \right)^2 
{\rm Re} \left[\tau_*^{-A} \int^0_{-\infty}d\tau' {\tau'}^{5+A} e^{-ik_t \tau'} \right] ~,  \\
f^{(a)}_{W^3}
&=& e_{ij}^{(-\lambda_1)} 
\left[ \frac{1}{2} e_{jk}^{(-\lambda_2)} e_{ki}^{(-\lambda_3)} 
+ \frac{3}{4} e_{kl}^{(-\lambda_2)} e_{kl}^{(-\lambda_3)} \hat{k_2}_i \hat{k_3}_j 
\right. \nonumber \\
&&\qquad\quad \left. 
+ \frac{3}{4}  e_{ki}^{(-\lambda_2)} e_{jl}^{(-\lambda_3)} \hat{k_2}_l \hat{k_3}_k
- \frac{3}{2}  e_{ik}^{(-\lambda_2)} e_{kl}^{(-\lambda_3)} \hat{k_2}_l \hat{k_3}_j
\right] + 5 \ {\rm perms} , \\
f_{\widetilde{W}W^2}^{(r)} 
&=&  
8 \left(\frac{H}{M_{\rm pl}}\right)^6 \left( \frac{H}{\Lambda} \right)^2 
{\rm Im} 
\left[\tau_*^{-A} \int^0_{-\infty}d\tau' {\tau'}^{5+A} e^{-ik_t \tau'} \right] ~, \\
f_{\widetilde{W}W^2}^{(a)}
&=& i \eta^{ijk} 
\left[ e_{kq}^{(-\lambda_1)} 
\left\{- 3 e_{jm}^{(-\lambda_2)} e_{iq}^{(-\lambda_3)}
\hat{k_3}_m + e_{mi}^{(-\lambda_2)} e_{mq}^{(-\lambda_3)} \hat{k_3}_j \right\} \right. \nonumber \\
&&\qquad \left. 
+ e_{pj}^{(-\lambda_1)} e_{pm}^{(-\lambda_2)}  
\hat{k_1}_k \hat{k_2}_l
\left\{ e_{il}^{(-\lambda_3)} \hat{k_3}_m 
- e_{im}^{(-\lambda_3)}  \hat{k_3}_l \right\} \right] + 5 \ {\rm perms}~.
\end{eqnarray}
Here, 
$k_t \equiv \sum_{n=1}^3 k_n$,
$int = W^3 $ and $\widetilde{W}W^2$, 
``5 perms'' denotes the five symmetric
terms under the permutations of $(\hat{\mib{k_1}}, \lambda_1),
(\hat{\mib{k_2}}, \lambda_2)$, and $(\hat{\mib{k_3}}, \lambda_3)$. 
From the above expressions,
we find that the bispectra of the primordial gravitational wave induced from $W^3$
and $\widetilde{W}W^2$ are proportional to the real and imaginary parts of 
$\tau_*^{-A} \int_{-\infty}^0 d\tau' \tau'^{5+A} e^{- i k_t \tau'}$,
respectively. 
This difference comes from the number of $\gamma_{ij ,\alpha
\alpha}$ and $\dot{\gamma}_{ij, k}$. 
$H_{W^3}$ consists of the products of an odd number of the former terms 
and an even number of the latter terms. 
On the other hand, in
$H_{\widetilde{W}W^2}$, the situation is the opposite. 
Since the former and latter terms contain 
$\ddot{\gamma}_{dS} - k^2 \gamma_{dS} 
= (2 H \tau' / M_{\rm pl}) k^{3/2} e^{-ik\tau'}$ and $\dot{\gamma}_{dS} 
= i ( H \tau' / M_{\rm pl}) k^{1/2} e^{-ik\tau'}$,
respectively, the total numbers of $i$ are different in each time
integral. Hence, the contributions of the real and imaginary parts roll upside
down in $f_{W^3}^{(r)}$ and $f_{\widetilde{W}W^2}^{(r)}$. 
Since the time integral in the bispectra can be analytically evaluated as
\begin{eqnarray}
\tau_*^{-A} \int_{-\infty}^0 d\tau' 
\tau'^{5+A} e^{- i k_t \tau'} 
= \left[ \cos\left( \frac{\pi}{2}A \right) 
+ i \sin \left(\frac{\pi}{2}A\right) \right]\Gamma(6+A) 
k_t^{-6}(-k_t \tau_*)^{-A}~, \nonumber \\
\end{eqnarray}
$f_{W^3}^{(r)}$ and $f_{\widetilde{W}W^2}^{(r)}$ reduce to
\begin{eqnarray}
f^{(r)}_{W^3} 
&=& 8 \left(\frac{H}{M_{\rm pl}}\right)^6 
\left( \frac{H}{\Lambda} \right)^2 
\cos\left(\frac{\pi}{2}A\right) \Gamma(6+A) k_t^{-6}(-k_t \tau_*)^{-A}
~,  \label{eq:radial_w3} \\
f_{\widetilde{W}W^2}^{(r)} 
&=&  
8 \left(\frac{H}{M_{\rm pl}}\right)^6 
\left( \frac{H}{\Lambda} \right)^2 
\sin\left(\frac{\pi}{2}A\right) \Gamma(6+A) k_t^{-6}(-k_t \tau_*)^{-A}
~, \label{eq:radial_ww2}
\end{eqnarray}
where $\Gamma(x)$ is the Gamma function.

From this equation, we can see that
in the case of the time-independent coupling, which corresponds to the $A = 0$ case,
the bispectrum from $\widetilde{W}W^2$ vanishes.
This is consistent with a claim in
Ref.~\citen{Soda:2011am}.
\footnote{
In Ref.~\citen{Soda:2011am}, the authors have shown that
for $A=0$, the bispectrum from $\widetilde{W}W^2$ has a nonzero value
upward in the first order of the slow-roll parameter.} 
On the other hand, interestingly, if $A$ deviates
 from $0$, it is possible to realize the nonzero bispectrum
 induced from $\widetilde{W}W^2$ even in the exact
 de Sitter limit. Thus, we expect the signals from $\widetilde{W}W^2$
 without the slow-roll suppression, which can be comparable to those from $W^3$
 and become sufficiently large to observe in the CMB.

\subsection{Running  coupling constant} \label{subsec:coupling}

Here, we discuss how to realize $f \propto \tau^A$ within the framework of the standard slow-roll inflation. 
During the standard slow-roll inflation,
the equation of motion of the scalar field $\phi$, which has a potential $V$,
 is expressed as 
\begin{eqnarray}
\dot{\phi} \simeq \pm \sqrt{2 \epsilon_\phi} M_{\rm pl} \tau^{-1} ~,
\end{eqnarray}
where $\epsilon_\phi \equiv [\partial V / \partial \phi /
(3 M_{\rm pl} H^2)]^2 / 2 $ is a slow-roll parameter
for $\phi$, 
$+$ and $-$ signs are taken to be for $\partial V/ \partial \phi > 0$ and $\partial V/ \partial \phi < 0$, respectively, and we have assumed that $ aH = -1/\tau$. 
The solution of the above equation is given by 
\begin{eqnarray}
\phi = \phi_* \pm \sqrt{2 \epsilon_\phi} M_{\rm pl} \ln \left( \frac{\tau}{\tau_*} \right)~.
\end{eqnarray}
Hence, if we assume a dilaton-like coupling as 
$f \equiv e^{(\phi - \phi_*) / M}$, we have 
\begin{eqnarray}
f(\tau) = \left( \frac{\tau}{\tau_*} \right)^A ~, \ \ 
A = \pm \sqrt{2 \epsilon_\phi} \frac{M_{\rm pl}}{M}~, \label{eq:coupling_moduli}
\end{eqnarray} 
where $M$ is an arbitrary energy scale.
Let us take $\tau_*$ to be a time when
the scale of the present horizon of the Universe exits
the horizon during inflation, namely,
$ \left| \tau_* \right| = k_*^{-1} \sim 14 {\rm Gpc}$.
Then, the coupling $f$, which determines the amplitude of
the bispectrum of the primordial gravitational wave induced
from the Weyl cubic terms, is on the order of unity for the
current cosmological scales.
From Eq.~(\ref{eq:coupling_moduli}),
we have $A = \pm 1/2$ with $M = \sqrt{8\epsilon_\phi} M_{\rm pl}$. 
As seen in Eqs.~(\ref{eq:radial_w3}) and (\ref{eq:radial_ww2}), this leads
 to an interesting situation that the bispectra from $W^3$ and
 $\widetilde{W}W^2$ have a comparable magnitude as $f^{(r)}_{W^3} = \pm f^{(r)}_{\widetilde{W}W^2}$. 
 Hence, we can expect that in the CMB bispectrum, the signals from these
 terms are almost the same. 

In the next section, we demonstrate these through the explicit calculation of the CMB bispectra. 
 
\section{CMB parity-even and parity-odd bispectrum}\label{sec:CMB_bis}

In this section, following the calculation approach discussed in
Ref.~\citen{Shiraishi:2010kd}, we formulate the CMB bispectrum induced
from the non-Gaussianities of gravitons sourced by $W^3$ and $\widetilde{W}W^2$ terms discussed in the
previous section.

\subsection{Formulation}\label{subsec:formulation}

Conventionally, the CMB fluctuation is expanded with the spherical harmonics as 
\begin{eqnarray}
\frac{\Delta X (\hat{\mib{n}})}{X} = \sum_{\ell m} a_{X, \ell m} Y_{\ell
 m}(\hat{\mib{n}})~, \label{eq:cmb_anisotropy}
\end{eqnarray}
where $\hat{\mib{n}}$ is a unit vector pointing toward a line-of-sight direction, and $X$ means the intensity ($\equiv I$) and
the electric and magnetic polarization modes ($\equiv E, B$). 
By performing the line-of-sight integration, the coefficient, $a_{\ell m}$,
generated from the primordial fluctuation of gravitons, $\gamma^{(\pm 2)}$, is
given by \cite{Shiraishi:2010kd, Weinberg:2008zzc, Shiraishi:2010sm}
\begin{eqnarray}
a_{X, \ell m} &=& 4\pi (-i)^\ell \int_0^\infty \frac{k^2 dk}{(2\pi)^3}
 {\cal T}_{X,\ell}(k) \sum_{\lambda = {\pm 2}} 
\left(\frac{\lambda}{2}\right)^x \gamma_{\ell m}^{(\lambda)}(k)~, \\
\gamma_{\ell m}^{(\lambda)}(k)  &\equiv& \int d^2 \hat{\mib{k}} 
\gamma^{(\lambda)}(\mib{k}) 
{}_{-\lambda}Y^*_{\ell m}(\hat{\mib{k}})~, \label{eq:gamma_lm}
\end{eqnarray}
where $x$ discriminates the parity of three modes: $x = 0$ for $X = I,E$
and $x=1$ for $X = B$, and ${\cal T}_{X, \ell}$ is the time-integrated
transfer function of tensor modes as calculated in, e.g., Refs.~\citen{Zaldarriaga:1996xe, Hu:1997hp, Pritchard:2004qp}.
Using this expression, we can obtain the CMB bispectrum generated from
the primordial bispectrum of gravitons as 
\begin{eqnarray}
\Braket{\prod_{n=1}^3 a_{X_n, \ell_n m_n}} 
&=&  \left[\prod_{n=1}^3 4\pi (-i)^{\ell_n} \int \frac{k_n^2 dk_n}{(2\pi)^3}
 {\cal T}_{X_n,\ell_n}(k_n) \sum_{\lambda_n = \pm 2} 
\left( \frac{\lambda_n}{2} \right)^{x_n} \right] \nonumber \\
&&\times
\Braket{\prod_{n=1}^3 \gamma_{\ell_n m_n}^{(\lambda_n)}(k_n) }~. \label{eq:cmb_bis_form}
\end{eqnarray}

In order to derive an explicit form of this CMB bispectrum, at first,
 we need to express all the functions containing the angular dependence
on the wave number vectors with the spin spherical harmonics. 
Using the results of Appendix \ref{appen:pol_tens}, 
$f^{(a)}_{W^3}$ and $f^{(a)}_{\widetilde{W}W^2}$ can be calculated as 
\begin{eqnarray}
f_{W^3}^{(a)}
&=&  
\left( 8\pi \right)^{3/2} 
\sum_{L', L'' = 2, 3} \sum_{M, M', M''} 
\left(
  \begin{array}{ccc}
   2 & L' & L'' \\
  M & M' & M''
  \end{array}
 \right) 
\nonumber \\
&&\times 
{}_{\lambda_1}Y_{2 M}^*(\hat{\mib{k_1}}) 
{}_{\lambda_2}Y_{L' M'}^*(\hat{\mib{k_2}}) {}_{\lambda_3}Y_{L''
M''}^*(\hat{\mib{k_3}}) 
\nonumber \\
&&\times 
\left[- \frac{1}{20}\sqrt{\frac{7}{3}} \delta_{L', 2} \delta_{L'', 2} 
+ (-1)^{L'} I_{L' 1 2}^{\lambda_2 0 -\lambda_2} I_{L'' 1 2}^{\lambda_3 0
-\lambda_3}
 \right. \nonumber \\
&&\qquad \left. 
 \times 
\left(  
- \frac{\pi}{5}
\left\{
  \begin{array}{ccc}
   2 & L' & L'' \\
   2 & 1 & 1 
  \end{array}
 \right\} 
- \pi 
\left\{
  \begin{array}{ccc}
   2 & L' & L'' \\
   1 & 1 & 2 \\
   1 & 2 & 1
  \end{array}
 \right\} \right. \right. \nonumber \\
&&\qquad\quad \left. \left. 
+ 2\pi 
\left\{
  \begin{array}{ccc}
   2 & 1 & L' \\
   2 & 1 & 1 
  \end{array}
 \right\}
\left\{
  \begin{array}{ccc}
   2 & L' & L'' \\
   2 & 1 & 1 
  \end{array}
 \right\}
\right)
\right] + 5 \ {\rm perms}
~, \label{eq:fa_w3} \\
f_{\widetilde{W}W^2}^{(a)}
&=& \left( 8\pi \right)^{3/2}  
\sum_{L', L'' = 2, 3} \sum_{M, M', M''} 
\left(
  \begin{array}{ccc}
   2 & L' & L'' \\
  M & M' & M''
  \end{array}
 \right) \nonumber \\
&&\times
{}_{\lambda_1}Y_{2 M}^*(\hat{\mib{k_1}}) 
{}_{\lambda_2}Y_{L' M'}^*(\hat{\mib{k_2}}) {}_{\lambda_3}Y_{L''
M''}^*(\hat{\mib{k_3}}) (-1)^{L''} I_{L'' 1 2}^{\lambda_3 0 -\lambda_3} 
\nonumber \\
&& \times 
\left[
\delta_{L',2} 
\left( 3 \sqrt{\frac{2\pi}{5}} 
\left\{
  \begin{array}{ccc}
   2 & 2 & L'' \\
   1 & 2 & 1 
  \end{array}
 \right\}
- 2 \sqrt{2\pi} 
\left\{
  \begin{array}{ccc}
   2 & 2 & L'' \\
   1 & 1 & 1 \\
   1 & 1 & 2
  \end{array}
 \right\} \right)
\right. \nonumber \\
&&\quad \left.  
+ \frac{\lambda_1}{2} 
I_{L' 1 2}^{\lambda_2 0 -\lambda_2} 
\left( - \frac{4 \pi}{3} 
\left\{
  \begin{array}{ccc}
   2 & L' & L'' \\
   1 & 2 & 1 \\
   1 & 1 & 2
  \end{array}
\right\}
+ \frac{2\pi}{15} \sqrt{\frac{7}{3}} 
\left\{
  \begin{array}{ccc}
   2 & L' & L'' \\
   1 & 2 & 2    
  \end{array}
\right\}
\right)
\right] \nonumber \\
&&+ 5 \ {\rm perms} ~, \label{eq:fa_ww2}
\end{eqnarray}
where the $2 \times 3$ matrix of a bracket, and the $2 \times 3$ and $3
\times 3$ matrices of a curly bracket denote the Wigner-$3j, 6j$ and $9j$
symbols, respectively, and 
\begin{eqnarray}
I^{s_1 s_2 s_3}_{l_1 l_2 l_3} 
\equiv \sqrt{\frac{(2 l_1 + 1)(2 l_2 + 1)(2 l_3 + 1)}{4 \pi}}
\left(
  \begin{array}{ccc}
  l_1 & l_2 & l_3 \\
   s_1 & s_2 & s_3 
  \end{array}
 \right)~.
\end{eqnarray} 
The delta function is also expanded as
\begin{eqnarray}
\delta\left( \sum_{n=1}^3 {\mib{k_n}} \right) 
&=& 8 \int_0^\infty y^2 dy 
\left[ \prod_{n=1}^3 \sum_{L_n M_n} 
 (-1)^{L_n/2} j_{L_n}(k_n y) 
Y_{L_n M_n}^*(\hat{\mib{k_n}}) \right] 
\nonumber \\
&&\times 
I_{L_1 L_2 L_3}^{0 \ 0 \ 0}
 \left(
  \begin{array}{ccc}
  L_1 & L_2 & L_3 \\
  M_1 & M_2 & M_3 
  \end{array}
 \right)~.
\end{eqnarray}
Next, we integrate all the spin spherical harmonics over
$\hat{\mib{k_1}}, \hat{\mib{k_2}}$ and $\hat{\mib{k_3}}$ as
\begin{eqnarray}
\int d^2 \hat{\mib{k_1}} {}_{- \lambda_1}Y_{\ell_1 m_1}^* Y_{L_1 M_1}^* {}_{\lambda_1} Y_{2M}^* &=& I_{\ell_1 L_1 2}^{\lambda_1 0 -\lambda_1}
\left(
  \begin{array}{ccc}
  \ell_1 & L_1 & 2 \\
  m_1 & M_1 & M 
  \end{array}
 \right) ~, \\
\int d^2 \hat{\mib{k_2}} {}_{- \lambda_2}Y_{\ell_2 m_2}^* Y_{L_2 M_2}^* {}_{\lambda_2} Y_{L' M'}^* &=& I_{\ell_2 L_2 L'}^{\lambda_2 0 -\lambda_2}
\left(
  \begin{array}{ccc}
  \ell_2 & L_2 & L' \\
  m_2 & M_2 & M' 
  \end{array}
 \right) ~, \\
\int d^2 \hat{\mib{k_3}} {}_{- \lambda_3}Y_{\ell_3 m_3}^* Y_{L_3 M_3}^* {}_{\lambda_3} Y_{L'' M''}^* &=& I_{\ell_3 L_3 L''}^{\lambda_3 0 -\lambda_3}
\left(
  \begin{array}{ccc}
  \ell_3 & L_3 & L'' \\
  m_3 & M_3 & M'' 
  \end{array}
 \right) ~.
\end{eqnarray}
Through the summation over the azimuthal
quantum numbers, the product of the above five Wigner-$3j$ symbols is expressed with
the Wigner-$9j$ symbols as
\begin{eqnarray}
&& \sum_{\substack{M_1 M_2 M_3 \\ M M' M''}}
\left(
  \begin{array}{ccc}
  L_1 &  L_2 & L_3 \\
   M_1 & M_2 & M_3
  \end{array}
 \right)
\left(
  \begin{array}{ccc}
  2 &  L' & L'' \\
   M & M' & M''
  \end{array}
 \right) 
\nonumber \\
&&\qquad \times
\left(
  \begin{array}{ccc}
  \ell_1 &  L_1 & 2 \\
   m_1 & M_1 & M
  \end{array}
 \right)
\left(
  \begin{array}{ccc}
  \ell_2 &  L_2 & L' \\
   m_2 & M_2 & M'
  \end{array}
 \right)
\left(
  \begin{array}{ccc}
  \ell_3 &  L_3 & L'' \\
   m_3 & M_3 & M''
  \end{array}
 \right) \nonumber \\
&& \qquad\qquad\qquad = 
\left(
  \begin{array}{ccc}
  \ell_1 & \ell_2 & \ell_3 \\
   m_1 & m_2 & m_3
  \end{array}
 \right)
\left\{
  \begin{array}{ccc}
  \ell_1 & \ell_2 & \ell_3 \\
   L_1 & L_2 & L_3 \\
   2 & L' & L'' \\
  \end{array}
 \right\}~.
\end{eqnarray} 
Finally, performing the summation over the helicities, namely,
$\lambda_1, \lambda_2$ and $\lambda_3$, as 
\begin{eqnarray}
\sum_{\lambda = \pm 2} 
\left(\frac{\lambda}{2}\right)^{x}
I_{\ell L 2}^{\lambda 0 -\lambda} 
&=&
\begin{cases}
2 I_{\ell L 2}^{2 0 -2}~,  & 
( \ell + L + x = {\rm even} ) \\
0~, & ( \ell + L + x = {\rm odd} )
\end{cases} \\
\sum_{\lambda = \pm 2} 
\left(\frac{\lambda}{2}\right)^{x}
I_{\ell L L'}^{\lambda 0 -\lambda}
I_{L' 1 2}^{\lambda 0 -\lambda}
&=&
\begin{cases}
2 I_{\ell L L'}^{2 0 -2} I_{L' 1 2}^{2 0 -2} ~, & 
( \ell + L + x = {\rm odd} ) \\
0 ~, & ( \ell + L + x = {\rm even} ) 
\end{cases} \\
\sum_{\lambda = \pm 2} 
\left(\frac{\lambda}{2}\right)^{x+1}
I_{\ell L 2}^{\lambda 0 -\lambda} 
&=&
\begin{cases}
2 I_{\ell L 2}^{2 0 -2}~,  & 
( \ell + L + x = {\rm odd} ) \\
0~, & ( \ell + L + x = {\rm even} )
\end{cases} 
\end{eqnarray}
and considering the selection rules of the Wigner symbols \cite{Shiraishi:2010kd}, we derive the CMB bispectrum generated from the non-Gaussianity of gravitons induced by $W^3$ as
\begin{eqnarray}
&& \Braket{\prod_{n=1}^3 a_{X_n, \ell_n m_n}}_{W^3}
= \left(
  \begin{array}{ccc}
  \ell_1 & \ell_2 & \ell_3 \\
   m_1 & m_2 & m_3
  \end{array}
 \right)
\int_0^\infty y^2 dy
\sum_{L_1 L_2 L_3} (-1)^{\frac{L_1 + L_2 + L_3}{2}} I_{L_1 L_2 L_3}^{0~0~0}
\nonumber \\
&&\qquad \times 
\left[\prod_{n=1}^3 \frac{2}{\pi} (-i)^{\ell_n} \int k_n^2 dk_n
 {\cal T}_{X_n,\ell_n}(k_n) j_{L_n}(k_n y)\right] f_{W^3}^{(r)}(k_1,
k_2, k_3) 
\nonumber \\
&&\qquad \times 
\left(8 \pi \right)^{3/2} 
\sum_{L', L'' = 2, 3} 8 I_{\ell_1 L_1 2}^{2 0 -2} I_{\ell_2 L_2 L'}^{2 0 -2} 
I_{\ell_3 L_3 L''}^{2 0 -2} 
\left\{
  \begin{array}{ccc}
  \ell_1 & \ell_2 & \ell_3 \\
   L_1 & L_2 & L_3 \\
   2 & L' & L'' \\
  \end{array}
 \right\}
\nonumber \\
&&\qquad \times  
\left[- \frac{1}{20}\sqrt{\frac{7}{3}} \delta_{L', 2} \delta_{L'', 2}
\left( \prod_{n=1}^3 {\cal D}_{L_n, \ell_n, x_n}^{(e)} \right) 
\right. \nonumber \\ 
&&\qquad\quad \left.  + (-1)^{L'} I_{L' 1 2}^{2 0 -2} I_{L'' 1 2}^{2 0 -2} 
{\cal D}_{L_1, \ell_1, x_1}^{(e)} {\cal D}_{L_2, \ell_2, x_2}^{(o)} 
{\cal D}_{L_3, \ell_3, x_3}^{(o)} \right. \nonumber \\
&&\qquad\qquad \left. \times
\left( - \frac{\pi}{5} 
\left\{
  \begin{array}{ccc}
   2 & L' & L'' \\
   2 & 1 & 1 
  \end{array}
 \right\} 
- \pi 
\left\{
  \begin{array}{ccc}
   2 & L' & L'' \\
   1 & 1 & 2 \\
   1 & 2 & 1
  \end{array}
 \right\} \right. \right. \nonumber \\
&&\qquad\qquad\quad \left. \left.
+ 2\pi 
\left\{
  \begin{array}{ccc}
   2 & 1 & L' \\
   2 & 1 & 1 
  \end{array}
 \right\}
\left\{
  \begin{array}{ccc}
   2 & L' & L'' \\
   2 & 1 & 1 
  \end{array}
 \right\}
\right)
\right]
+ 5 \ {\rm perms} ~, \label{eq:cmb_bis_w3}
\end{eqnarray}
and $\widetilde{W}W^2$ as 
\begin{eqnarray}
&&\Braket{\prod_{n=1}^3 a_{X_n, \ell_n m_n}}_{\widetilde{W}W^2}
= \left(
  \begin{array}{ccc}
  \ell_1 & \ell_2 & \ell_3 \\
   m_1 & m_2 & m_3
  \end{array}
 \right)
\int_0^\infty y^2 dy
\sum_{L_1 L_2 L_3} (-1)^{\frac{L_1 + L_2 + L_3}{2}} I_{L_1 L_2 L_3}^{0~0~0}
\nonumber \\
&&\qquad \times 
\left[\prod_{n=1}^3 \frac{2}{\pi} (-i)^{\ell_n} \int k_n^2 dk_n
 {\cal T}_{X_n,\ell_n}(k_n) j_{L_n}(k_n y)\right] f_{\widetilde{W}W^2}^{(r)}(k_1,
k_2, k_3) 
\nonumber \\
&&\qquad \times 
\left( 8\pi \right)^{3/2} 
\sum_{L', L'' = 2, 3} 8 I_{\ell_1 L_1 2}^{2 0 -2} I_{\ell_2 L_2 L'}^{2 0 -2}
I_{\ell_3 L_3 L''}^{2 0 -2} 
\left\{
  \begin{array}{ccc}
  \ell_1 & \ell_2 & \ell_3 \\
   L_1 & L_2 & L_3 \\
   2 & L' & L'' \\
  \end{array}
 \right\}
(-1)^{L''} I_{L'' 1 2}^{2 0 -2} 
\nonumber \\
&&\qquad \times 
\left[
\delta_{L',2}
{\cal D}_{L_1, \ell_1, x_1}^{(e)}
{\cal D}_{L_2, \ell_2, x_2}^{(e)} {\cal D}_{L_3, \ell_3, x_3}^{(o)}
\right.  \nonumber \\
&&\qquad\qquad 
\left. \times \left( 
3 \sqrt{\frac{2\pi}{5}}
\left\{
  \begin{array}{ccc}
   2 & 2 & L'' \\
   1 & 2 & 1 
  \end{array}
 \right\}
- 2 \sqrt{2\pi}  
\left\{
  \begin{array}{ccc}
   2 & 2 & L'' \\
   1 & 1 & 1 \\
   1 & 1 & 2
  \end{array}
 \right\} \right) \right. \nonumber \\
&&\quad \left.  
\qquad + I_{L' 1 2}^{2 0 -2} 
\left(\prod_{n=1}^3 {\cal D}_{L_n, \ell_n, x_n}^{(o)} \right) 
\right. \nonumber \\
&&\qquad\qquad \left. \times
\left( - \frac{4 \pi}{3} 
\left\{
  \begin{array}{ccc}
   2 & L' & L'' \\
   1 & 2 & 1 \\
   1 & 1 & 2
  \end{array}
\right\}
+ \frac{2\pi}{15} \sqrt{\frac{7}{3}} 
\left\{
  \begin{array}{ccc}
   2 & L' & L'' \\
   1 & 2 & 2    
  \end{array}
\right\} 
\right) \right] 
+ 5 \ {\rm perms}~. \nonumber \\
\label{eq:cmb_bis_ww2}
\end{eqnarray}
Here, ``5 perms'' denotes the five symmetric
terms under the permutations of $(\ell_1, m_1, x_1)$, $(\ell_2, m_2,
x_2)$, and $(\ell_3, m_3, x_3)$, and 
we introduce the filter functions as 
\begin{eqnarray}
{\cal D}^{(e)}_{L, \ell, x} 
&\equiv& (\delta_{L, \ell - 2} + \delta_{L, \ell} + \delta_{L, \ell + 2}) 
\delta_{x, 0}
\nonumber \\
&& + (\delta_{L, |\ell - 3|} + \delta_{L, \ell - 1} + \delta_{L, \ell + 1} + \delta_{L, \ell + 3}) \delta_{x,1} ~, \\
{\cal D}^{(o)}_{L, \ell, x} 
&\equiv& (\delta_{L, \ell - 2} + \delta_{L, \ell} + \delta_{L, \ell + 2}) 
\delta_{x, 1} \nonumber \\ 
&& + ( \delta_{L, |\ell - 3|} + \delta_{L, \ell - 1} +
\delta_{L, \ell + 1} + \delta_{L, \ell + 3} ) \delta_{x,
0} ~,
\end{eqnarray}
where the superscripts $(e)$ and $(o)$ denote $L + \ell + x = {\rm even}$
and $= {\rm odd}$, respectively. 
From Eqs.~(\ref{eq:cmb_bis_w3}) and (\ref{eq:cmb_bis_ww2}), we can see
that the azimuthal quantum numbers $m_1, m_2$, and $m_3$ are confined only in a Wigner-$3j$ symbol as 
$\left(
  \begin{array}{ccc}
  \ell_1 & \ell_2 & \ell_3 \\
   m_1 & m_2 & m_3
  \end{array}
 \right)$.
This guarantees the rotational invariance of the CMB
bispectrum. Therefore, this bispectrum survives if the triangle
inequality is satisfied as $|\ell_1 - \ell_2| \leq \ell_3 \leq \ell_1 + \ell_2 $. 

Considering the products between the ${\cal D}$ functions in Eq.~(\ref{eq:cmb_bis_w3}) and the selection rules as $\sum_{n=1}^3 L_n = {\rm even}$, we can notice that the CMB bispectrum from $W^3$ does not vanish only for 
\begin{eqnarray}
\sum_{n=1}^3 (\ell_n + x_n) = {\rm even} ~.
\end{eqnarray} 
Therefore, $W^3$ contributes the $III, IIE, IEE, IBB, EEE$, and $EBB$
spectra for $\sum_{n=1}^3 \ell_n = {\rm even}$ and the $IIB, IEB, EEB$,
and $BBB$ spectra for $\sum_{n=1}^3
\ell_n = {\rm odd}$. This property can arise from any sources keeping the
parity invariance such as $W^3$. 
On the other hand, in the same manner, we understand that the CMB bispectrum from $\widetilde{W}W^2$ survives only for 
\begin{eqnarray}
\sum_{n=1}^3 (\ell_n + x_n) = {\rm odd} ~.
\end{eqnarray} 
By these constraints, we find that in reverse, $\widetilde{W}W^2$
generates the $IIB, IEB, EEB$, and $BBB$ spectra for $\sum_{n=1}^3 \ell_n =
{\rm even}$ and the $III, IIE, IEE, IBB, EEE$, and $EBB$ spectra for $\sum_{n=1}^3 \ell_n = {\rm odd}$. This
is a characteristic signature of the parity violation as mentioned in
Refs.~\citen{Okamoto:2002ik} and \citen{Kamionkowski:2010rb}. Hence, if we analyze the 
information of the CMB bispectrum not only for $\sum_{n=1}^3 \ell_n =
{\rm even}$ but also for $\sum_{n=1}^3 \ell_n = {\rm odd}$, it may be
possible to check the parity violation at the level of the
three-point correlation. 

The above discussion about the multipole configurations of the CMB
bispectra can be easily understood only if one consider the parity
transformation of the CMB intensity and polarization fields in the real space (\ref{eq:cmb_anisotropy}). 
The $III$, $IIE$, $IEE$, $IBB$, $EEE$ and
 $EBB$ spectra from $W^3$, and the $IIB, IEB, EEB$, and $BBB$ spectra
 from $\widetilde{W}W^2$ have even parity, namely, 
\begin{eqnarray} 
\Braket{\prod_{i=1}^3 \frac{\Delta X_i (\hat{\mib{n_i}})}{X_i}} 
= \Braket{\prod_{i=1}^3 \frac{\Delta X_i (-\hat{\mib{n_i}})}{X_i}}~. 
\end{eqnarray}
Then, from the multipole expansion
(\ref{eq:cmb_anisotropy}) and its parity flip version as 
\begin{eqnarray}
\frac{\Delta X (- \hat{\mib{n}})}{X} = \sum_{\ell m} a_{X, \ell m} Y_{\ell
 m}(- \hat{\mib{n}}) = \sum_{\ell m} (-1)^\ell a_{X, \ell m} Y_{\ell
 m}(\hat{\mib{n}})~,
\end{eqnarray}
one can notice that $\sum_{n=1}^3 \ell_n = {\rm even}$ must be satisfied. 
On the other hand, since the $IIB, IEB, EEB$, and $BBB$ spectra
 from $W^3$, and the $III$, $IIE$, $IEE$, $IBB$, $EEE$, and
 $EBB$ spectra from $\widetilde{W}W^2$ have odd parity, namely, 
\begin{eqnarray} 
\Braket{\prod_{i=1}^3 \frac{\Delta X_i (\hat{\mib{n_i}})}{X_i}} 
= - \Braket{\prod_{i=1}^3 \frac{\Delta X_i (-\hat{\mib{n_i}})}{X_i}}~, 
\end{eqnarray}
one can obtain $\sum_{n=1}^3 \ell_n = {\rm odd}$.

In \S\ref{subsec:results}, we compute the CMB
bispectra (\ref{eq:cmb_bis_w3}) and (\ref{eq:cmb_bis_ww2}) 
when $A = \pm 1/2, 0, 1$, that is, the signals
from $W^3$ become as large as those from $\widetilde{W}W^2$ and either signals vanish.
  
\subsection{Evaluation of $f_{W^3}^{(r)}$ and $f_{\widetilde{W}W^2}^{(r)}$}

\begin{figure}[t]
  \begin{tabular}{cc}
    \begin{minipage}{0.5\hsize}
  \begin{center}
    \includegraphics[width=7.0cm,height=5.5cm,clip]{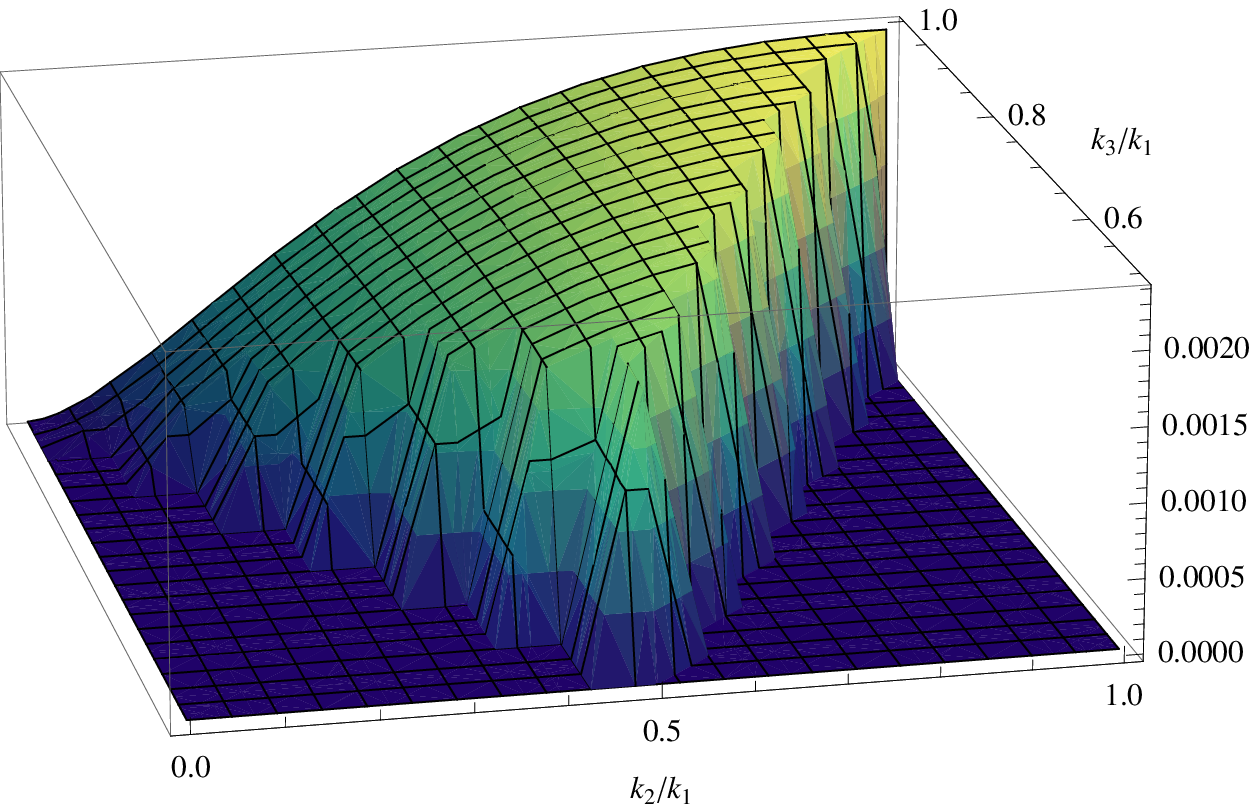}
  \end{center}
\end{minipage}
\begin{minipage}{0.5\hsize}
  \begin{center}
    \includegraphics[width=7.0cm,height=5.5cm,clip]{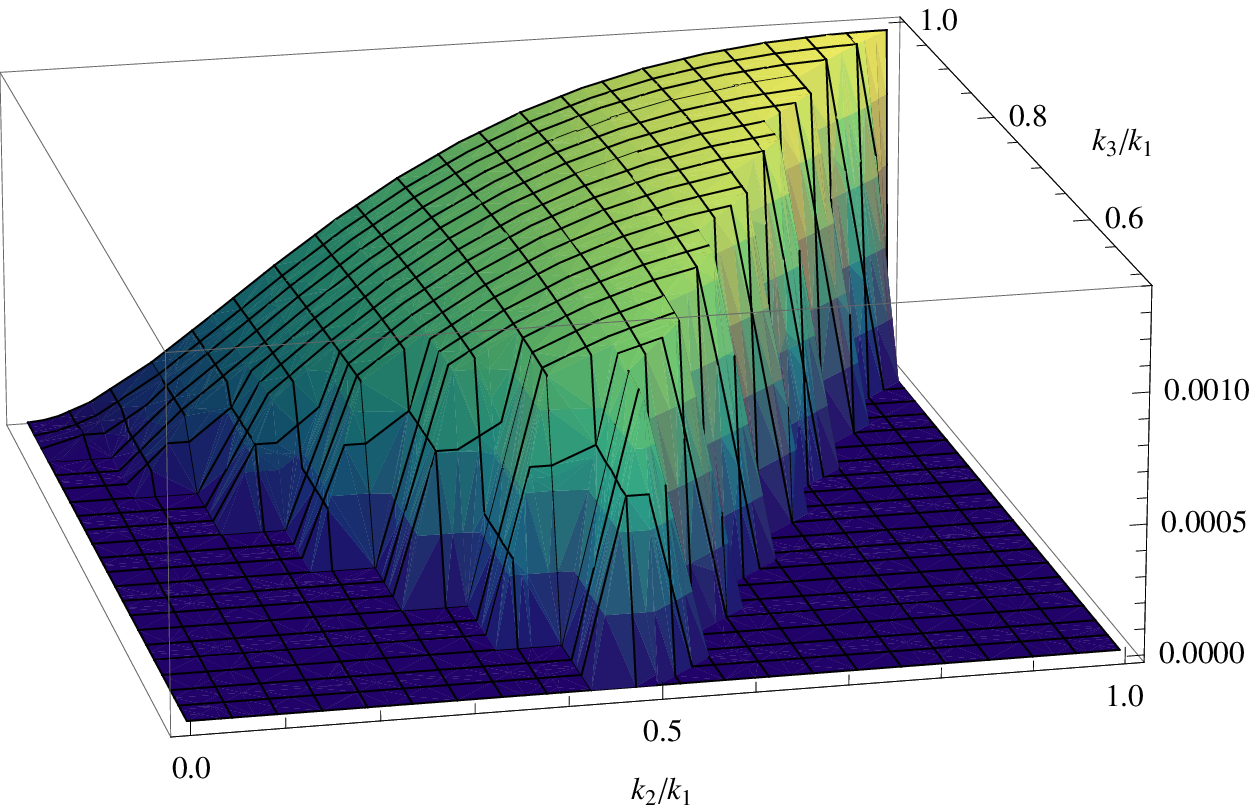}
  \end{center}
\end{minipage}
\end{tabular}
\\
  \begin{tabular}{cc}
    \begin{minipage}{0.5\hsize}
  \begin{center}
    \includegraphics[width=7.0cm,height=5.5cm,clip]{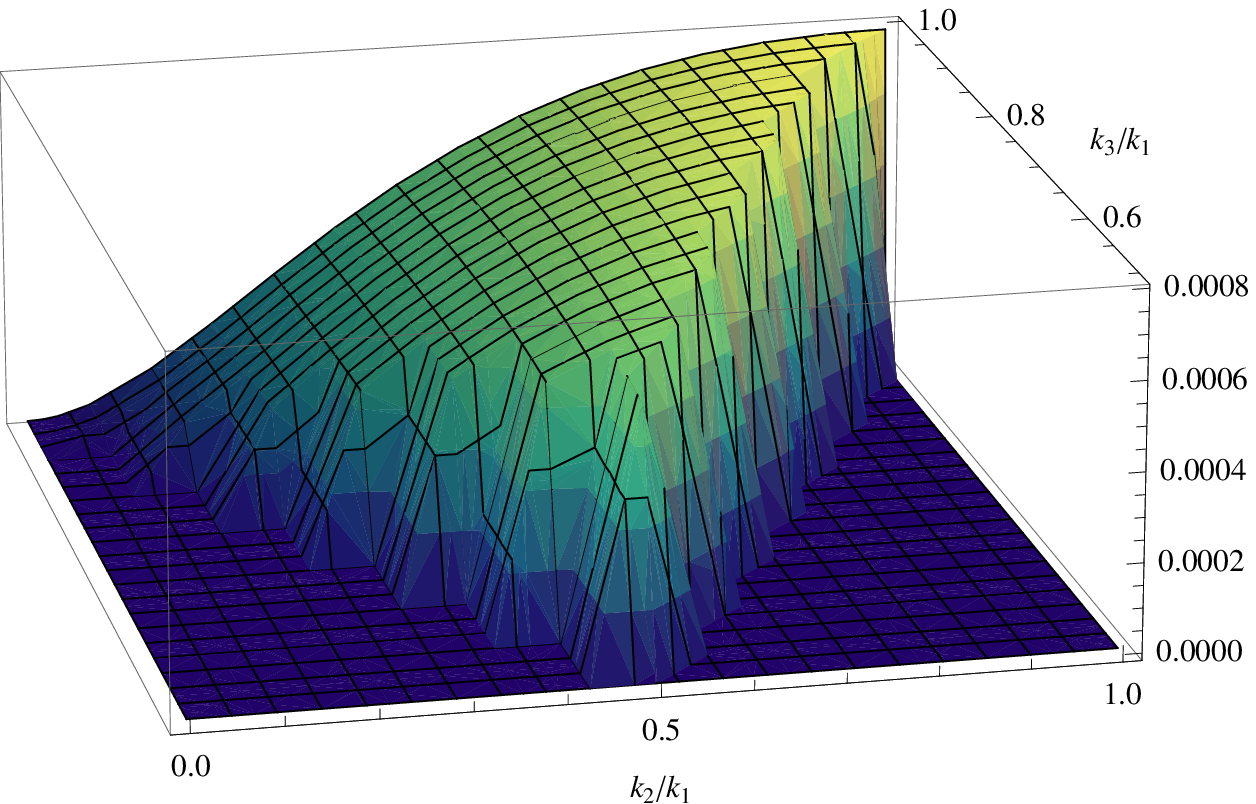}
  \end{center}
\end{minipage}
\begin{minipage}{0.5\hsize}
  \begin{center}
    \includegraphics[width=7.0cm,height=5.5cm,clip]{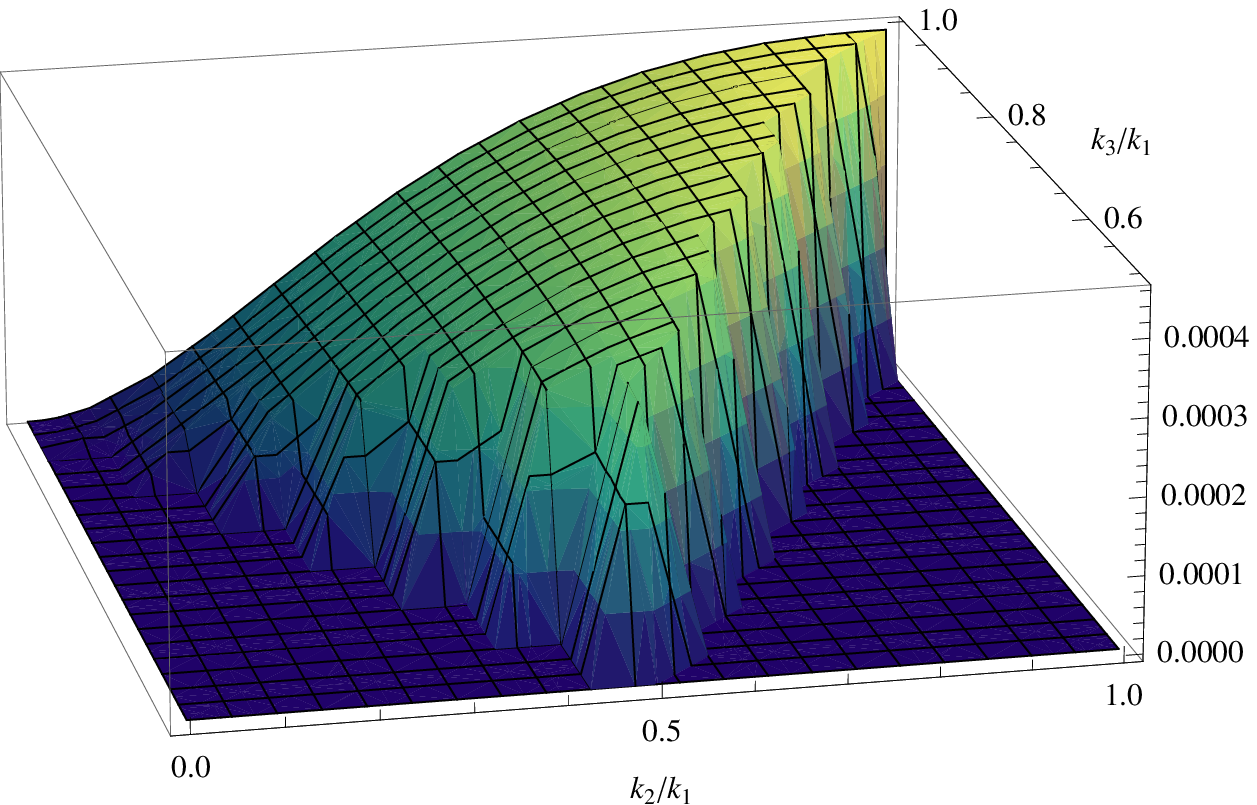}
  \end{center}
\end{minipage}
\end{tabular}
  \caption{Shape of $k_1^2 k_2^2 k_3^2 S_{A}$ for $A = -1/2$ (top left figure), $0$ (top right one), $1/2$ (bottom left one), and $1$ (bottom right one) as the function of $k_2 / k_1$ and $k_3 / k_1$.} \label{fig:S}
\end{figure}

Here, to compute the CMB bispectra (\ref{eq:cmb_bis_w3}) and
(\ref{eq:cmb_bis_ww2}) in finite time, we express the radial functions,
 $f^{(r)}_{W^3}$ and $f^{(r)}_{\widetilde{W}W^2}$, with some terms of the
 power of $k_1, k_2$, and $k_3$. 
Let us focus on the dependence on $k_1, k_2$, and $k_3$ in
Eqs.~(\ref{eq:radial_w3}) and (\ref{eq:radial_ww2}) as
\begin{eqnarray}
f_{W^3}^{(r)} \propto f_{\widetilde{W}W^2}^{(r)} &\propto& k_t^{-6}(-k_t \tau_*)^{-A} 
=  \frac{S_A(k_1, k_2, k_3)}{(k_1 k_2 k_3)^{A/3} (-\tau_*)^{A}}~, 
\end{eqnarray}
where we define $S_A$ to satisfy $S_A \propto k^{-6}$ as
\begin{eqnarray}
S_A(k_1, k_2, k_3) &\equiv&
\frac{(k_1 k_2 k_3)^{A/3}}{k_t^{6+A}}~.
\end{eqnarray}
In Fig.~\ref{fig:S}, we plot $S_{A}$ for $A = -1/2, 0, 1/2$, and $1$. From this, we
notice that the shapes of $S_A$ are similar to the equilateral-type 
configuration as \cite{Creminelli:2005hu}
\begin{eqnarray}
S_{\rm eq}(k_1, k_2, k_3) 
&=& 6 \left(- \frac{1}{k_1^3 k_2^3} - \frac{1}{k_2^3 k_3^3} - \frac{1}{k_3^3 k_1^3} - \frac{2}{k_1^2 k_2^2 k_3^2} \right. \nonumber \\
&&\quad \left. 
+ \frac{1}{k_1 k_2^2 k_3^3} + \frac{1}{k_1 k_3^2 k_2^3}
+ \frac{1}{k_2 k_3^2 k_1^3} + \frac{1}{k_2 k_1^2 k_3^3}
+ \frac{1}{k_3 k_1^2 k_2^3} + \frac{1}{k_3 k_2^2 k_1^3}
 \right)~. \nonumber \\
\end{eqnarray}
To evaluate how a function $S$ is similar in shape to a function $S'$, we introduce a correlation function as \cite{Babich:2004gb, Senatore:2009gt}
\begin{eqnarray}
\cos(S \cdot S') \equiv \frac{S \cdot S'}
{(S \cdot S)^{1/2} (S' \cdot S')^{1/2} }~,
\end{eqnarray}
with
\begin{eqnarray}
S \cdot S' 
&\equiv& \sum_{\mib{k_i}} \frac{S(k_1, k_2, k_3) S'(k_1, k_2, k_3)}
{P(k_1) P(k_2) P(k_3)} \nonumber \\
&\propto& \int_0^1 d x_2 
\int_{1-x_2}^1 d x_3 x_2^4 x_3^4 S(1, x_2, x_3) S'(1, x_2, x_3)~,
\end{eqnarray}
where the summation is performed over all $\mib{k_i}$, which form a triangle and $P(k) \propto k^{-3}$ denotes the power spectrum. 
This correlation function gets to 1 when $S = S'$. 
In our case, this is calculated as  
\begin{eqnarray}
\cos (S_{A} \cdot S_{\rm eq}) 
&\simeq& 
\begin{cases}
0.968 ~, & (A = -1/2) \\
0.970 ~, & (A = 0) \\
0.971 ~, & (A = 1/2) \\
0.972 ~, & (A = 1) \\
\end{cases}
\end{eqnarray}
that is, an approximation that $S_A$ is proportional to $S_{\rm eq}$
seems to be valid. Here, we also calculate the correlation functions with the local- and orthogonal-type non-Gaussianities \cite{Komatsu:2010fb} and conclude that these contributions are negligible. Thus, we determine the proportionality coefficient as 
\begin{eqnarray}
S_A &\simeq& 
\frac{S_A \cdot S_{\rm eq}}{S_{\rm eq} \cdot
S_{\rm eq}}  S_{\rm eq} 
= 
\begin{cases}
4.40 \times 10^{-4}S_{\rm eq} ~, & (A = -1 / 2) \\
2.50 \times 10^{-4}S_{\rm eq} ~, & (A = 0) \\
1.42 \times 10^{-4}S_{\rm eq} ~, & (A = 1 / 2) \\
8.09 \times 10^{-5}S_{\rm eq} ~. & (A = 1)
\end{cases}
\end{eqnarray}
Substituting this into Eqs.~(\ref{eq:radial_w3}) and (\ref{eq:radial_ww2}), we obtain reasonable formulae of the radial functions
for $A = 1/2$ as
\begin{eqnarray}
f_{W^3}^{(r)} &=& f_{\widetilde{W}W^2}^{(r)}
 \nonumber \\
&\simeq&  \left( \frac{\pi^2}{2} r A_S \right)^4 \left( \frac{M_{\rm pl}}{\Lambda} \right)^2 
\frac{10395}{8} \sqrt{\frac{\pi}{2}} \times
\frac{1.42 \times 10^{-4} S_{\rm eq}}{ (-\tau_*)^{1/2} (k_1 k_2 k_3)^{1/6} }
~, \label{eq:fr_+0.5} 
\end{eqnarray}
and for $A = - 1/2$ as
\begin{eqnarray}
f_{W^3}^{(r)} &=& - f_{\widetilde{W}W^2}^{(r)}
 \nonumber \\ 
&\simeq&  \left( \frac{\pi^2}{2} r A_S \right)^4 \left( \frac{M_{\rm pl}}{\Lambda} \right)^2 
\frac{945}{4} \sqrt{\frac{\pi}{2}} 
\nonumber \\
&&
\times 4.40 \times 10^{-4} (-\tau_*)^{1/2} (k_1 k_2 k_3)^{1/6} S_{\rm eq}~. 
\label{eq:fr_-0.5}
\end{eqnarray}
Here, we also use 
\begin{eqnarray}
\left( \frac{H}{M_{\rm pl}} \right)^2 = \frac{\pi^2}{2} r A_S~,
\end{eqnarray}
where $A_S$ is the amplitude of primordial curvature perturbations and $r$ is the tensor-to-scalar ratio \cite{Komatsu:2010fb, Shiraishi:2010kd}. 
For $A = 0$, the signals from $\widetilde{W}W^2$ disappear as
$f^{(r)}_{\widetilde{W}W^2} = 0$ and the finite radial function of $W^3$
is given by 
\begin{eqnarray}
f^{(r)}_{W^3} \simeq  
\left( \frac{\pi^2}{2} r A_S \right)^4 \left( \frac{M_{\rm pl}}{\Lambda} \right)^2 960 \times 2.50 \times 10^{-4}S_{\rm eq}~. \label{eq:fr_0}
\end{eqnarray}
In contrast, for $A = 1$, since $f^{(r)}_{W^3} = 0$, we have only the parity-violating contribution from $\widetilde{W}W^2$ as 
\begin{eqnarray}
f^{(r)}_{\widetilde{W}W^2} \simeq 
\left( \frac{\pi^2}{2} r A_S \right)^4 
\left( \frac{M_{\rm pl}}{\Lambda} \right)^2 5760 \times
\frac{8.09 \times 10^{-5} S_{\rm eq}}{(- \tau_*) (k_1 k_2 k_3)^{1/3}}~. \label{eq:fr_1}
\end{eqnarray} 

\subsection{Results} \label{subsec:results}

On the basis of the analytical formulae (\ref{eq:cmb_bis_w3}), (\ref{eq:cmb_bis_ww2}), (\ref{eq:fr_+0.5}), (\ref{eq:fr_-0.5}), (\ref{eq:fr_0}) and (\ref{eq:fr_1}), we compute the CMB bispectra 
from $W^3$ and $\widetilde{W}W^2$ for $A = - 1/2, 0 , 1/2$, and $1$.
 Then, we modify the Boltzmann Code for Anisotropies in the
Microwave Background (CAMB) \cite{Lewis:1999bs,Lewis:2004ef}. In
calculating the Wigner symbols, we use the Common
Mathematical Library SLATEC \cite{slatec} and some analytic formulae described in Ref.~\citen{Shiraishi:2010kd}.   
From the dependence of the radial functions $f_{W^3}^{(r)}$ and $f_{\widetilde{W}W^2}^{(r)}$ on the wave numbers, we can see that 
the shapes of the CMB
bispectra from $W^3$ and $\widetilde{W}W^2$ are similar to the
equilateral-type configuration. Then, the significant signals arise from
multipoles satisfying $\ell_1 \simeq \ell_2 \simeq \ell_3$. We confirm this by calculating the CMB bispectrum for several $\ell$'s. Hence, in the
following discussion, we give the discussion with the spectra for $\ell_1 \simeq \ell_2 \simeq \ell_3$. However, we do not focus on the spectra from $\sum_{n=1}^3 \ell_n = {\rm odd}$ for $\ell_1 = \ell_2 = \ell_3$ because these vanish due to the asymmetric nature.
 
\begin{figure}[!t]
  \begin{center}
    \includegraphics[height=17.6cm,clip]{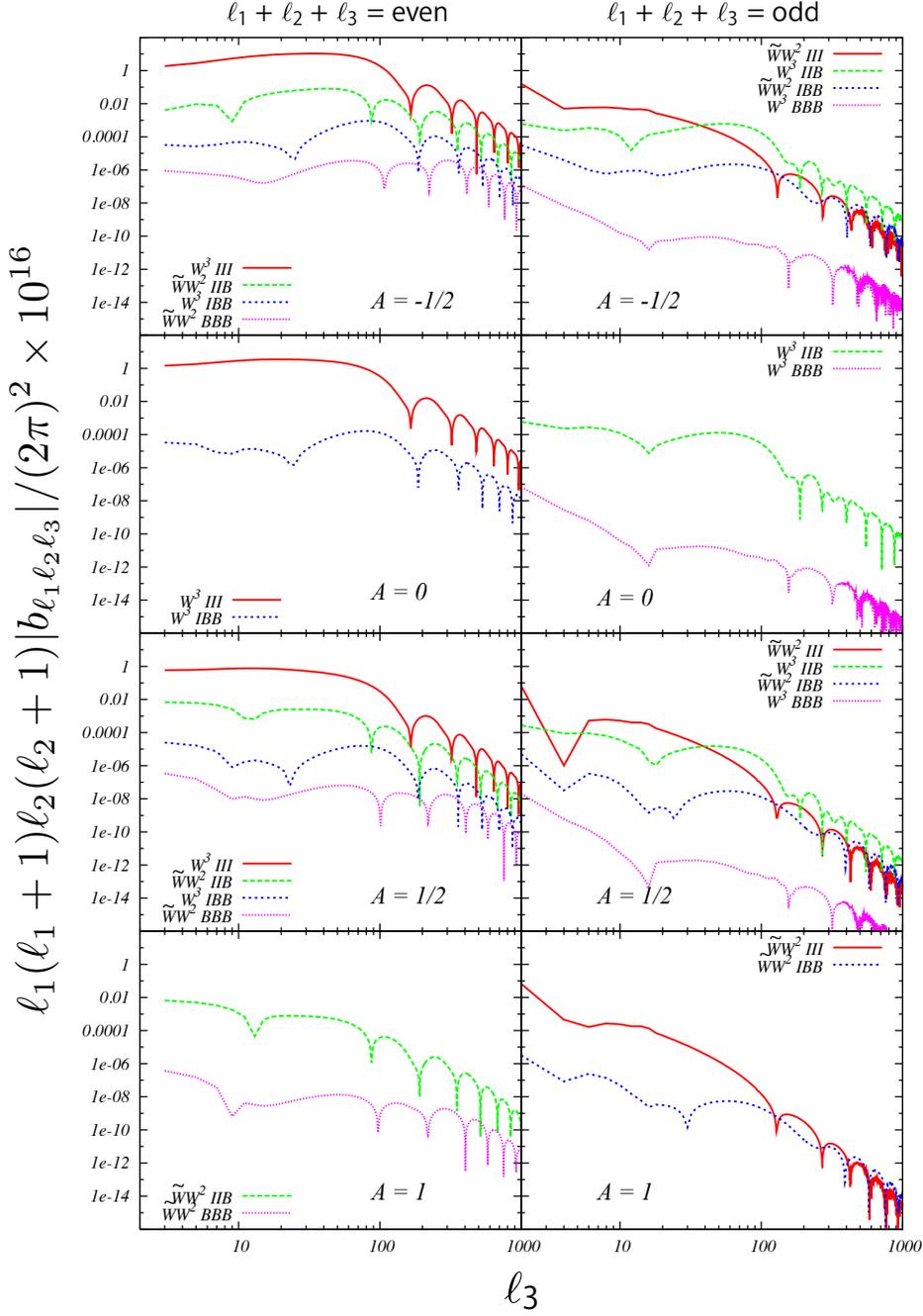}
  \end{center}
  \caption{Absolute values of the CMB $III, IIB, IBB$, and $BBB$ spectra induced by $W^3$ and $\widetilde{W}W^2$ for $A
 = -1/2, 0, 1/2$, and $1$. We set that three multipoles have identical values as $\ell_1 - 2 = \ell_2 - 1 = \ell_3$. 
The left figures show the spectra not
 vanishing for $\sum_{n=1}^3 \ell_n = {\rm even}$ (parity-even mode)
 and the right ones present the spectra for $\sum_{n=1}^3 \ell_n = {\rm
 odd}$ (parity-odd mode). Here, we fix the parameters as $ \Lambda =
 3 \times 10^6 {\rm GeV}, r = 0.1$, and $\tau_* = -k_*^{-1} = - 14 {\rm Gpc}$, and other
 cosmological parameters are fixed as the mean values limited from the
 WMAP $7$-yr data\cite{Komatsu:2010fb}.} \label{fig:TTT_A_difl}
\end{figure}

In Fig.~\ref{fig:TTT_A_difl}, we present the reduced CMB $III, IIB,
IBB$, and $BBB$ spectra given by
\begin{eqnarray}
b_{X_1 X_2 X_3, \ell_1 \ell_2 \ell_3}
 &=&  ( G_{\ell_1 \ell_2 \ell_3} )^{-1} \sum_{m_1 m_2 m_3} 
\left(
  \begin{array}{ccc}
  \ell_1 & \ell_2 & \ell_3 \\
   m_1 & m_2 & m_3
  \end{array}
 \right)
\Braket{\prod_{n=1}^3 a_{X_n, \ell_n m_n}} ~, \nonumber \\ 
\end{eqnarray}
for $\ell_1 - 2  = \ell_2 - 1 = \ell_3$. Here, the $G$ symbol is defined by 
\cite{Kamionkowski:2010rb}, 
\footnote{The conventional expression of the CMB-reduced bispectrum as 
\begin{eqnarray}
b_{X_1 X_2 X_3, \ell_1 \ell_2 \ell_3} \equiv (I_{\ell_1 \ell_2 \ell_3}^{0~0~0})^{-1} 
\sum_{m_1 m_2 m_3} 
\left(
  \begin{array}{ccc}
  \ell_1 & \ell_2 & \ell_3 \\
   m_1 & m_2 & m_3
  \end{array}
 \right)
\Braket{\prod_{n=1}^3 a_{X_n, \ell_n m_n}} 
\end{eqnarray}
breaks down for $\sum_{n=1}^{3} \ell_n = {\rm odd}$ due to the
divergence behavior of $(I_{\ell_1 \ell_2 \ell_3}^{0~0~0})^{-1}$. Here,
replacing the $I$ symbol with the $G$ symbol, this problem is avoided. Of course, for $\sum_{n=1}^{3}
\ell_n = {\rm even}$,  $G_{\ell_1 \ell_2 \ell_3}$ is identical to $I_{\ell_1 \ell_2
\ell_3}^{0~0~0}$.
}
\begin{eqnarray}
G_{\ell_1 \ell_2 \ell_3} 
&\equiv& \frac{2 \sqrt{\ell_3 (\ell_3 + 1) \ell_2 (\ell_2 +
1)}}{\ell_1(\ell_1 + 1) - \ell_2 (\ell_2 + 1) - \ell_3 (\ell_3 + 1)}
\nonumber \\
&&\times
\sqrt{\frac{\prod_{n=1}^3 (2 \ell_n + 1)}{4 \pi}}
\left(
  \begin{array}{ccc}
  \ell_1 & \ell_2 & \ell_3 \\
   0 & -1 & 1
  \end{array}
 \right)~.
\end{eqnarray}
At first, from this figure, we can confirm that there are similar
features of the CMB power spectrum of tensor modes
\cite{Pritchard:2004qp, Baskaran:2006qs}. In the $III$
spectra, the dominant signals are located in $\ell < 100$ due to the
enhancement of the integrated Sachs-Wolfe effect. On the other
hand, since the fluctuation of polarizations is mainly produced through
the Thomson scattering at around the recombination and reionization epoch,
the $BBB$ spectra have two peaks for $\ell < 10$ and $\ell \sim 100$,
respectively. The cross-correlated bispectra between $I$ and $B$ modes
seem to contain both these effects. These features back up the consistency of our calculation.
 
The curves in Fig.~\ref{fig:TTT_A_difl} denote the spectra for $A =
-1/2, 0, 1/2$, and $1$, respectively. 
We notice that the spectra for large $A$
become red compared with those for small $A$. 
The difference in tilt of $\ell$ between these spectra is just one corresponding to the difference in $A$.  
The curves of the left and right figures obey $\sum_{n=1}^3 \ell_n
= {\rm even}$ and $= {\rm odd}$, respectively. 
As mentioned in \S\ref{subsec:formulation}, we stress again that in the
$\ell$ configuration where the bispectrum from $W^3$ vanishes,
the bispectrum from $\widetilde{W}W^2$ survives, and vice versa
for each correlation. This is because the parities of these terms
are opposite each other. For example, this predicts a nonzero $III$ spectrum not only for $\sum_{n=1}^3 \ell_n = {\rm even}$
due to $W^3$ but also for $\sum_{n=1}^3 \ell_n = {\rm odd}$ due
to $\widetilde{W}W^2$. 

We can also see that each bispectrum induced by $W^3$ has a different shape from that induced by $\widetilde{W}W^2$ corresponding to the difference in the
primordial bispectra. 
Regardless of this, the overall amplitudes of the spectra for $A = \pm 1/2$ are almost identical. 
However, if we consider $A$ deviating from these values, the
balance between the contributions of $W^3$ and $\widetilde{W}W^2$ breaks. 
For example, if $-1/2 < A < 1/2$, the contribution of $W^3$
dominates. Assuming the time-independent coupling, namely, $A = 0$, since
$f^{(r)}_{\widetilde{W}W^2} = 0$, the CMB bispectra are generated only
from $W^3$. Thus, we will never observe the parity violation of gravitons
in the CMB bispectrum. On the other hand, when $-3/2 < A < -1/2$ or $1/2
< A < 3/2$, the contribution of $\widetilde{W}W^2$ dominates. In an
extreme case, if $A = {\rm odd}$, since $f^{(r)}_{W^3} = 0$, the CMB
bispectra arise only from $\widetilde{W}W^2$ and violate the parity invariance. Then, the information of the signals under $\sum_{n=1}^3 \ell_n = {\rm odd}$ will become more important in the analysis of the $III$ spectrum.

\begin{figure}[t]
  \begin{center}
    \includegraphics[height=6cm,clip]{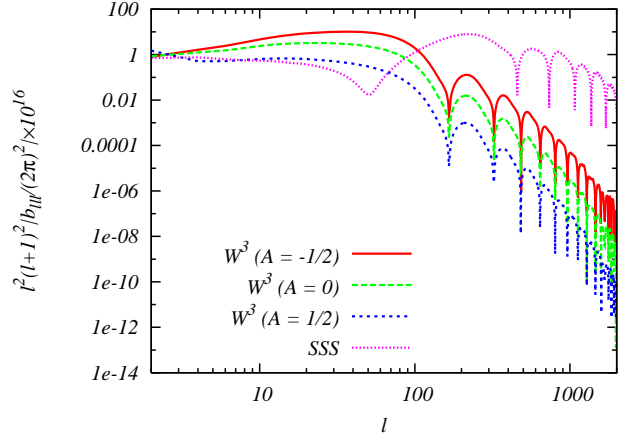}
  \end{center}
  \caption{(color online) Absolute value of the CMB $III$ spectra generated from $W^3$
 for $A = -1/2$ (red solid line), $0$ (green dashed one), and $1/2$ (blue dotted one) described in Fig.~\ref{fig:TTT_A_difl}, and generated from the equilateral-type
 non-Gaussianity given by Eq.~(\ref{eq:cmb_bis_scal}) with $f_{\rm
 NL}^{\rm eq} = 300$ (magenta dot-dashed one). 
 We set that three multipoles have identical values as $\ell_1 = \ell_2
 = \ell_3 \equiv \ell$. Here, we fix the parameters as the same values
 mentioned in Fig.~\ref{fig:TTT_A_difl}.} \label{fig:TTT_A_III_even_samel}
\end{figure}

In Fig.~\ref{fig:TTT_A_III_even_samel}, we focus on the $III$
spectra from $W^3$ for $\ell_1 = \ell_2 = \ell_3 \equiv \ell$ to compare these with the $III$ spectrum generated from the equilateral-type
non-Gaussianity of curvature perturbations given by 
\begin{eqnarray}
b_{III, \ell_1 \ell_2 \ell_3}^{(SSS)}  
&=& \int_0^\infty y^2 dy 
\left[ \prod_{n=1}^3 \frac{2}{\pi} \int_0^\infty k_n^2 dk_n 
{\cal T}_{I,\ell_n}^{(S)}(k_n) j_{\ell_n}(k_n y) \right] 
\nonumber \\
&&\times \frac{3}{5} f_{\rm NL}^{\rm eq} (2\pi^2 A_S)^2 
S_{\rm eq}(k_1, k_2, k_3) ~, \label{eq:cmb_bis_scal}
\end{eqnarray}
where $f_{\rm NL}^{\rm eq}$ is the nonlinearity
parameter of the equilateral non-Gaussianity and ${\cal T}_{I,
\ell}^{(S)}$ is the transfer function of scalar mode \cite{Zaldarriaga:1996xe, Hu:1997hp}. Note that these
three spectra vanish for $\sum_{n=1}^3 \ell_n = {\rm odd}$. 
From this figure,
we can estimate the typical amplitude of the $III$ spectra from $W^3$ at large scale as 
\begin{eqnarray}
|b_{\ell \ell \ell}| \sim \ell^{-4} \times 3.2 \times 10^{-2}
\left(\frac{{\rm GeV}}{\Lambda}\right)^2 \left(\frac{r}{0.1}\right)^4 ~.
\end{eqnarray}
This equation also seems to be applicable to the $III$ spectra from
$\widetilde{W}W^2$. 
On the other hand, the CMB bispectrum generated from the
equilateral-type non-Gaussianity on a large scale is evaluated with $f_{\rm NL}^{\rm eq}$ as
\begin{eqnarray}
|b_{\ell \ell \ell}| 
\sim \ell^{-4} \times 4 \times 10^{-15}  
\left|\frac {f_{\rm NL}^{\rm eq}}{300}\right| ~.
\end{eqnarray} 
From these estimations and ideal upper bounds on $f_{\rm NL}^{\rm eq}$
estimated only from the cosmic variance for $\ell < 100$
\cite{Creminelli:2005hu, Creminelli:2006rz, Smith:2006ud}, namely
$f_{\rm NL}^{\rm eq} \lesssim 300$ and $r \sim 0.1$, we find a rough
limit: $\Lambda \gtrsim 3 \times 10^6 {\rm GeV}$. Here, we use only the signals for $\sum_{n=1}^3 \ell_n = {\rm even}$ due to the comparison with the parity-conserving bispectrum from scalar-mode non-Gaussianity. Of course, to estimate more precisely, we will have to calculate the signal-to-noise ratio with the information of $\sum_{n=1}^3 \ell_n = {\rm odd}$ \cite{Kamionkowski:2010rb}. 

\section{Summary and discussion}

In this paper, we have studied the CMB bispectrum generated from the graviton
non-Gaussianity induced by the parity-even and parity-odd Weyl cubic terms,
namely, $W^3$ and $\widetilde{W}W^2$,  which have a dilaton-like coupling
depending on the conformal time as $f \propto \tau^A$. Through the
calculation based on the in-in formalism, we have found that the primordial
non-Gaussianities from $\widetilde{W}W^2$ can have a magnitude comparable to that from $W^3$ even in the exact de Sitter
space-time. 

Using the explicit formulae of the primordial bispectrum, we
have derived the CMB bispectra of the intensity ($I$) and polarization ($E,B$)
modes. Then, we have confirmed that, owing to the difference in the 
transformation under parity,
the spectra from $W^3$ vanish in the $\ell$ space where those
from $\widetilde{W}W^2$ survive and vice versa. For example, 
owing to the parity-violating $\widetilde{W}W^2$ term,
the $III$ spectrum can be produced not only for $\sum_{n=1}^3 \ell_n = {\rm even}$ but
also for $\sum_{n=1}^3 \ell_n = {\rm odd}$, and the $IIB$ spectrum can
also be produced
for $\sum_{n=1}^3 \ell_n = {\rm even}$.
These signals are powerful lines of evidence the parity violation in the non-Gaussian level; hence, to
reanalyze the observational data for $\sum_{n=1}^3 \ell_n = {\rm odd}$
is meaningful work. 

When $A = -1/2, 0, 1/2$, and $1$, we have obtained reasonable numerical
results of the CMB bispectra from the parity-conserving $W^3$ and 
the parity-violating $\widetilde{W}W^2$. For $A = \pm 1/2$, we have
found that the spectra from $W^3$ and $\widetilde{W}W^2$ have almost the
same magnitudes 
even though these have a small difference in the shapes. 
In contrast, if $A = 0$ and $1$, we have confirmed that the signals from
$\widetilde{W}W^2$ and $W^3$ vanish, respectively. In the latter case,
we will observe only the parity-violating signals in the CMB bispectra
generated from the Weyl cubic terms. 
We have also found that
the shape of the non-Gaussianity from such Weyl cubic terms
is quite similar to the equilateral-type
non-Gaussianity of curvature perturbations. 
In comparison with the $III$ spectrum 
generated from the equilateral-type non-Gaussianity,
we have found that if $r = 0.1$, $\Lambda \gtrsim 3 \times 10^6 {\rm GeV}$ 
corresponds approximately to $f_{\rm NL}^{\rm eq} \lesssim 300$. 

Strictly speaking, to obtain the bound on the scale $\Lambda$, we need to
calculate the signal-to-noise ratio with the information of not only
$\sum_{n=1}^3 \ell_n = {\rm even}$ but also $\sum_{n=1}^3 \ell_n = {\rm
odd}$ for each $A$ by the application of Ref.~\citen{Kamionkowski:2010rb}. 
This will be discussed in the future. 

\section*{Acknowledgements}
We would like to thank Juan M. Maldacena, Jiro Soda, Hideo Kodama, and
Masato Nozawa for useful comments. 
This work was supported in part by a Grant-in-Aid for JSPS Research under
 Grant No.~22-7477 (M. S.), JSPS Grant-in-Aid for Scientific
Research under Grant No.~22340056 (S. Y.), Grant-in-Aid for Scientific Research on Priority Areas No.~467 ``Probing the Dark Energy through an Extremely Wide and Deep Survey with
Subaru Telescope'', and Grant-in-Aid for Nagoya University
 Global COE Program ``Quest for Fundamental Principles in the Universe:
 from Particles to the Solar System and the Cosmos'', from the Ministry
 of Education, Culture, Sports, Science and Technology of Japan. 
We also acknowledge the Kobayashi-Maskawa Institute for the Origin of
Particles and the Universe, Nagoya University, for providing computing
resources useful in conducting the research reported in this paper. 

\appendix
\section{Calculation of $f_{W^3}^{(a)}$ and $f_{\widetilde{W}W^2}^{(a)}$} \label{appen:pol_tens}

Here, we calculate each product between the wave number vectors and the
polarization tensors of $f_{W^3}^{(a)}$ and $f_{\widetilde{W}W^2}^{(a)}$
mentioned in \S\ref{subsec:formulation} \cite{Shiraishi:2010kd, Shiraishi:2011fi}. 

We set the polarization tensor defined in Refs.~\citen{Shiraishi:2010kd}
and \citen{Weinberg:2008zzc} as 
\begin{eqnarray}
e^{(\lambda)}_{ab}(\hat{\mib{k}}) \equiv \frac{1}{\sqrt{2}}
\left(\hat{\theta}_a(\hat{\mib{k}}) + i \frac{\lambda}{2}
 \hat{\phi}_a(\hat{\mib{k}}) \right) 
\left(\hat{\theta}_b(\hat{\mib{k}}) + i \frac{\lambda}{2}
 \hat{\phi}_b(\hat{\mib{k}}) \right)~, \label{eq:pol_tens_def}
\end{eqnarray}
with 
\begin{eqnarray}
\hat{\mib{k}} \equiv 
\left(
  \begin{array}{c}
    \sin\theta \cos\phi \\
    \sin\theta \sin\phi \\
    \cos\theta
  \end{array}
\right)~, \ \ 
\hat{\mib{\theta}} \equiv 
\left(
  \begin{array}{c}
    \cos\theta \cos\phi  \\
    \cos\theta \sin\phi \\
    -\sin \theta
  \end{array}
\right)~, \ \ 
\hat{\mib{\phi}} \equiv 
\left(
  \begin{array}{c}
    -\sin \phi  \\
    \cos \phi \\
    0
  \end{array}
\right)~.
\end{eqnarray} 
Here, $\lambda = \pm 2$ denotes the helicity of the gravitational wave. 
Of course, this polarization tensor obeys the relations ({\ref{eq:pol_tens_rel}}). 
According to Ref.~\citen{Shiraishi:2010kd}, a unit vector and a polarization tensor (\ref{eq:pol_tens_def}) are expanded with the spin spherical harmonics respectively, as
\begin{eqnarray}
\hat{k}_a &=& \sum_m \alpha_a^{m} Y_{1 m}(\hat{\mib{k}})~, \\ 
e^{(\lambda)}_{ab} (\hat{\mib{k}})
&=& \frac{3}{\sqrt{2 \pi}}  
\sum_{M m_a m_b} {}_{-\lambda}Y_{2 M}^*(\hat{\mib{k}}) 
\alpha^{m_a}_{a} \alpha^{m_b}_b 
\left(
  \begin{array}{ccc}
  2 & 1 &  1\\
  M & m_a & m_b 
  \end{array}
\right)~,
\end{eqnarray}
with 
\begin{eqnarray}
\mib{\alpha}^m &\equiv& \sqrt{\frac{2 \pi}{3}}
 \left(
  \begin{array}{ccc}
   -m (\delta_{m,1} + \delta_{m,-1}) \\
   i~ (\delta_{m,1} + \delta_{m,-1}) \\
   \sqrt{2} \delta_{m,0}
  \end{array}
\right)~. \label{eq:arbitrary_vec}
\end{eqnarray}
Then, the scalar product of $\mib{\alpha}^m$ is given by
\begin{eqnarray}
\alpha_a^m \alpha_a^{m'} = \frac{4 \pi}{3} (-1)^m \delta_{m,-m'}~.
\end{eqnarray}
Using these relations, the first term of $f_{W^3}^{(a)}$ is written as
\begin{eqnarray}
e_{ij}^{(-\lambda_1)} e_{jk}^{(-\lambda_2)} e_{ki}^{(-\lambda_3)}
&=& - ( 8\pi )^{3/2} 
\sum_{M, M', M''} {}_{\lambda_1}Y_{2 M}^*(\hat{\mib{k_1}}) 
{}_{\lambda_2}Y_{2 M'}^*(\hat{\mib{k_2}}) {}_{\lambda_3}Y_{2 M''}^*(\hat{\mib{k_3}})
\nonumber \\
&&\times 
\frac{1}{10} \sqrt{\frac{7}{3}}
\left(
  \begin{array}{ccc}
   2 & 2 & 2 \\
  M & M' & M''
  \end{array}
 \right)~,
\end{eqnarray} 
where the summation of three Wigner symbols included in the polarization tensors with respect to azimuthal quantum numbers is performed using a formula: 
\begin{eqnarray}
&& \sum_{m_4 m_5 m_6} (-1)^{\sum_{i=4}^6 l_i - m_i}
\left(
  \begin{array}{ccc}
  l_5 & l_1 & l_6 \\
  m_5 & -m_1 & -m_6 
  \end{array}
 \right) \nonumber \\
&&\qquad \times
\left(
  \begin{array}{ccc}
  l_6 & l_2 & l_4 \\
  m_6 & -m_2 & -m_4 
  \end{array}
 \right)
\left(
  \begin{array}{ccc}
  l_4 & l_3 & l_5 \\
  m_4 & -m_3 & -m_5 
  \end{array}
 \right) \nonumber \\
 &&\qquad\qquad\qquad = \left(
  \begin{array}{ccc}
  l_1 & l_2 & l_3 \\
  m_1 & m_2 & m_3 
  \end{array}
 \right) 
\left\{
  \begin{array}{ccc}
  l_1 & l_2 & l_3 \\
  l_4 & l_5 & l_6 
  \end{array}
 \right\}~.
 \end{eqnarray}
In the same manner, we can obtain the other terms of $f_{W^3}^{(a)}$ as
\begin{eqnarray}
&&e_{ij}^{(-\lambda_1)} e_{kl}^{(-\lambda_2)} e_{kl}^{(-\lambda_3)} \hat{k_2}_i \hat{k_3}_j \nonumber \\
&&\qquad = - (8\pi)^{3/2} 
\sum_{L', L'' = 2, 3} \sum_{M, M', M''} {}_{\lambda_1}Y_{2 M}^*(\hat{\mib{k_1}}) 
{}_{\lambda_2}Y_{L' M'}^*(\hat{\mib{k_2}}) {}_{\lambda_3}Y_{L'' M''}^*(\hat{\mib{k_3}}) 
\nonumber \\
&&\qquad\quad \times 
\frac{4\pi}{15}(-1)^{L'} I_{L' 1 2}^{\lambda_2 0 -\lambda_2} I_{L'' 1 2}^{\lambda_3 0 -\lambda_3}
\left(
  \begin{array}{ccc}
   2 & L' & L'' \\
  M & M' & M''
  \end{array}
 \right)
\left\{
  \begin{array}{ccc}
   2 & L' & L'' \\
  2 & 1 & 1
  \end{array}
 \right\}
~, \\
&&e_{ij}^{(-\lambda_1)} e_{ki}^{(-\lambda_2)} e_{jl}^{(-\lambda_3)} \hat{k_2}_l \hat{k_3}_k \nonumber \\
&&\qquad = -( 8\pi )^{3/2} 
\sum_{L', L'' = 2, 3} \sum_{M, M', M''} 
{}_{\lambda_1}Y_{2 M}^*(\hat{\mib{k_1}}) 
{}_{\lambda_2}Y_{L' M'}^*(\hat{\mib{k_2}}) {}_{\lambda_3}Y_{L'' M''}^*(\hat{\mib{k_3}}) 
\nonumber \\
&&\qquad\quad \times 
\frac{4\pi}{3} (-1)^{L'} I_{L' 1 2}^{\lambda_2 0 -\lambda_2} I_{L'' 1 2}^{\lambda_3 0 -\lambda_3}
\left(
  \begin{array}{ccc}
   2 & L' & L'' \\
  M & M' & M''
  \end{array}
 \right)
\left\{
  \begin{array}{ccc}
   2 & L' & L'' \\
   1 & 1 & 2 \\
   1 & 2 & 1
  \end{array}
 \right\}~, \nonumber \\
\\
&& e_{ij}^{(-\lambda_1)} e_{ik}^{(-\lambda_2)} e_{kl}^{(-\lambda_3)} \hat{k_2}_l \hat{k_3}_j \nonumber \\
&&\qquad= - (8\pi)^{3/2} 
\sum_{L', L'' = 2, 3} \sum_{M, M', M''} {}_{\lambda_1}Y_{2 M}^*(\hat{\mib{k_1}}) 
{}_{\lambda_2}Y_{L' M'}^*(\hat{\mib{k_2}}) {}_{\lambda_3}Y_{L'' M''}^*(\hat{\mib{k_3}}) 
\nonumber \\
&&\qquad\quad \times 
\frac{4\pi}{3} (-1)^{L'} I_{L' 1 2}^{\lambda_2 0 -\lambda_2} I_{L'' 1 2}^{\lambda_3 0 -\lambda_3}
\left(
  \begin{array}{ccc}
   2 & L' & L'' \\
  M & M' & M''
  \end{array}
 \right) \nonumber \\
&&\qquad\quad \times
\left\{
  \begin{array}{ccc}
   2 & 1 & L' \\
   2 & 1 & 1 
  \end{array}
 \right\}
\left\{
  \begin{array}{ccc}
   2 & L' & L'' \\
   2 & 1 & 1 
  \end{array}
 \right\}
~.
\end{eqnarray}
Here, in addition to the above relations, we use the product formula: 
\begin{eqnarray}
\prod_{i=1}^2 {}_{s_i} Y_{l_i m_i}(\hat{\mib{k}})
 = \sum_{l_3 m_3 s_3} {}_{s_3} Y^*_{l_3 m_3}(\hat{\mib{k}}) 
I^{-s_1 -s_2
-s_3}_{l_1~l_2~l_3} 
\left(
  \begin{array}{ccc}
  l_1 & l_2 & l_3 \\
  m_1 & m_2 & m_3
  \end{array}
 \right)~, \label{eq:product_sYlm} 
\end{eqnarray}
with 
\begin{eqnarray}
I^{s_1 s_2 s_3}_{l_1 l_2 l_3} 
\equiv \sqrt{\frac{(2 l_1 + 1)(2 l_2 + 1)(2 l_3 + 1)}{4 \pi}}
\left(
  \begin{array}{ccc}
  l_1 & l_2 & l_3 \\
   s_1 & s_2 & s_3 
  \end{array}
 \right)~,
\end{eqnarray}
and the summation rules of the Wigner symbols:
\begin{eqnarray}
&& (2 l_3 + 1) \sum_{m_1 m_2} 
\left(
  \begin{array}{ccc}
  l_1 & l_2 & l_3 \\
  m_1 & m_2 & m_3
  \end{array}
 \right)
\left(
  \begin{array}{ccc}
  l_1 & l_2 & l'_3 \\
  m_1 & m_2 & m'_3
  \end{array}
 \right) 
= \delta_{l_3, l_3'} \delta_{m_3, m'_3}~, \label{eq:Wig_3j_lllmmm} \\
&& \sum_{\substack{m_4 m_5 m_6 \\ m_7 m_8 m_9}} 
\left(
  \begin{array}{ccc}
  l_4 & l_5 & l_6 \\
  m_4 & m_5 & m_6 
  \end{array}
 \right)
\left(
  \begin{array}{ccc}
  l_7 & l_8 & l_9 \\
  m_7 & m_8 & m_9 
  \end{array}
 \right) \nonumber \\
&&\qquad \times 
\left(
  \begin{array}{ccc}
  l_4 & l_7 & l_1 \\
  m_4 & m_7 & m_1 
  \end{array}
 \right)
\left(
  \begin{array}{ccc}
  l_5 & l_8 & l_2 \\
  m_5 & m_8 & m_2
  \end{array}
 \right)
\left(
  \begin{array}{ccc}
  l_6 & l_9 & l_3 \\
  m_6 & m_9 & m_3 
  \end{array}
 \right) \nonumber \\ 
&& \qquad\qquad\qquad = \left(
  \begin{array}{ccc}
  l_1 & l_2 & l_3 \\
  m_1 & m_2 & m_3
  \end{array}
 \right)
\left\{
  \begin{array}{ccc}
  l_1 & l_2 & l_3 \\
  l_4 & l_5 & l_6 \\
  l_7 & l_8 & l_9 
  \end{array}
 \right\}~. \label{eq:sum_9j_3j}
\end{eqnarray}

In the calculation of $f_{\widetilde{W}W^2}^{(a)}$, we also need to
consider the dependence of the tensor contractions on $\eta^{ijk}$. Making use of the relation:
\begin{eqnarray}
\eta^{abc} \alpha^{m_a}_a \alpha^{m_b}_b \alpha^{m_c}_c 
&=& -i \left(\frac{4\pi}{3}\right)^{3/2} \sqrt{6}
\left(
  \begin{array}{ccc}
   1 & 1 & 1 \\
  m_a & m_b & m_c
  \end{array}
 \right) ~,
\end{eqnarray}
the first two terms of $f_{\widetilde{W}W^2}^{(a)}$ reduce to 
\begin{eqnarray}
&&i \eta^{ijk} e_{kq}^{(-\lambda_1)} e_{jm}^{(-\lambda_2)} e_{iq}^{(-\lambda_3)} \hat{k_3}_m \nonumber \\
&&\qquad= - ( 8\pi)^{3/2} 
\sum_{L'' = 2, 3} \sum_{M, M', M''} {}_{\lambda_1}Y_{2 M}^*(\hat{\mib{k_1}}) 
{}_{\lambda_2}Y_{2 M'}^*(\hat{\mib{k_2}}) {}_{\lambda_3}Y_{L'' M''}^*(\hat{\mib{k_3}}) \nonumber \\
&&\qquad\quad \times \sqrt{\frac{2\pi}{5}} (-1)^{L''}  
I_{L'' 1 2}^{\lambda_3 0 -\lambda_3} 
\left(
  \begin{array}{ccc}
   2 & 2 & L'' \\
  M & M' & M''
  \end{array}
 \right)
\left\{
  \begin{array}{ccc}
   2 & 2 & L'' \\
  1 & 2 & 1
  \end{array}
 \right\} ~,
\\
&& i \eta^{ijk} e_{kq}^{(-\lambda_1)} e_{mi}^{(-\lambda_2)}
e_{mq}^{(-\lambda_3)} \hat{k_3}_j \nonumber \\
&&\qquad = -( 8\pi )^{3/2} 
\sum_{L'' = 2, 3} \sum_{M, M', M''} {}_{\lambda_1}Y_{2 M}^*(\hat{\mib{k_1}}) 
{}_{\lambda_2}Y_{2 M'}^*(\hat{\mib{k_2}}) {}_{\lambda_3}Y_{L'' M''}^*(\hat{\mib{k_3}}) \nonumber \\
&&\qquad\quad \times 2\sqrt{2\pi} (-1)^{L''} I_{L'' 1 2}^{\lambda_3 0 -\lambda_3} 
\left(
  \begin{array}{ccc}
   2 & 2 & L'' \\
  M & M' & M''
  \end{array}
 \right)
\left\{
  \begin{array}{ccc}
   2 & 2 & L'' \\
  1 & 1 & 1 \\
  1 & 1 & 2
  \end{array}
 \right\}~.
\end{eqnarray}
For the other terms, by using the relation 
\begin{eqnarray}
\eta^{abc} \hat{k}_a e_{bd}^{(\lambda)}(\hat{\mib{k}}) 
&=& -\frac{\lambda}{2} i e_{cd}^{(\lambda)}(\hat{\mib{k}}) ~, \label{eq:eigen}
\end{eqnarray}
 we have 
\begin{eqnarray}
&& i \eta^{ijk} e_{pj}^{(-\lambda_1)} e_{pm}^{(-\lambda_2)}  
\hat{k_1}_k \hat{k_2}_l e_{il}^{(-\lambda_3)} \hat{k_3}_m \nonumber \\ 
&&\qquad= - \frac{\lambda_1}{2} (8\pi)^{3/2} \sum_{L', L'' = 2, 3} \sum_{M, M', M''} {}_{\lambda_1}Y_{2 M}^*(\hat{\mib{k_1}}) 
{}_{\lambda_2}Y_{L' M'}^*(\hat{\mib{k_2}}) {}_{\lambda_3}Y_{L'' M''}^*(\hat{\mib{k_3}}) \nonumber \\
&&\qquad\quad \times \frac{4\pi}{3}(-1)^{L''} I_{L' 1 2}^{\lambda_2 0 -\lambda_2}
I_{L'' 1 2}^{\lambda_3 0 -\lambda_3} 
\left(
  \begin{array}{ccc}
   2 & L' & L'' \\
  M & M' & M''
  \end{array}
 \right)
\left\{
  \begin{array}{ccc}
   2 & L' & L'' \\
  1 & 2 & 1 \\
  1 & 1 & 2
  \end{array}
 \right\} ~, \\
&& i \eta^{ijk} e_{pj}^{(-\lambda_1)} e_{pm}^{(-\lambda_2)}  
\hat{k_1}_k \hat{k_2}_l e_{im}^{(-\lambda_3)}  \hat{k_3}_l  \nonumber \\
&&\qquad=  - \frac{\lambda_1}{2} (8\pi)^{3/2}
 \sum_{L', L'' = 2, 3} \sum_{M, M', M''} {}_{\lambda_1}Y_{2 M}^*(\hat{\mib{k_1}}) 
{}_{\lambda_2}Y_{L' M'}^*(\hat{\mib{k_2}}) {}_{\lambda_3}Y_{L'' M''}^*(\hat{\mib{k_3}}) \nonumber \\
&&\qquad\quad \times \frac{2\pi}{15} \sqrt{\frac{7}{3}} (-1)^{L''} 
I_{L' 1 2}^{\lambda_2 0 -\lambda_2}
I_{L'' 1 2}^{\lambda_3 0 -\lambda_3} 
\left(
  \begin{array}{ccc}
   2 & L' & L'' \\
  M & M' & M''
  \end{array}
 \right)
\left\{
  \begin{array}{ccc}
   2 & L' & L'' \\
  1 & 2 & 2 
  \end{array}
 \right\}~. \nonumber \\
\end{eqnarray}


\end{document}